\documentclass{article}
\usepackage[a4paper,top=2cm,bottom=2cm,left=2cm,right=2cm]{geometry}
\usepackage{authblk}
  \usepackage{paralist}
  \usepackage{graphics}
  \usepackage{epsfig} 
\usepackage{graphicx}  \usepackage{epstopdf}
 \usepackage[colorlinks=true]{hyperref}
\hypersetup{urlcolor=blue, citecolor=red}
\usepackage{amsmath} 
\usepackage{amsfonts}
\usepackage[bbgreekl]{mathbbol}
\DeclareSymbolFontAlphabet{\mathbbm}{bbold}
\DeclareSymbolFontAlphabet{\mathbb}{AMSb}
\usepackage{amssymb}
\usepackage{amsthm}
\usepackage{mathrsfs}
\usepackage{stackrel,graphicx}
\usepackage{enumitem}
\usepackage{cases}
\usepackage{amsmath}
\usepackage{booktabs}
\usepackage{braket}
\usepackage[strict]{changepage}
\usepackage{subfigure}
\usepackage[small,justification=justified]{caption}
\usepackage{titlesec} 
\usepackage{multicol}
\usepackage{hyperref}
\usepackage{chemarr}

\usepackage{cite}

\usepackage{amssymb,amsmath,amsthm}
\usepackage{booktabs}
\usepackage{multirow}
\usepackage[table]{xcolor}
\usepackage{subfigure}
\usepackage{chemfig}
\usepackage{mhchem}

\newtheorem*{rem*}{Remark}
\theoremstyle{definition}

\newcommand{\eps}{\varepsilon}
\newcommand{\R}{\mathbb R}
\newcommand{\Sbb}{\mathbb S^{n-1}}

\providecommand{\keywords}[1]{\textbf{\textit{Keyworks --}} #1}

\newenvironment{sistem}
{\left\lbrace\begin{array}{@{}l@{}}}
{\end{array}\right.}

\begin{document}

\title{Mathematical modeling of glioma invasion: acid- and vasculature mediated go-or-grow dichotomy and the influence of tissue anisotropy}
\author[1]{Martina Conte\thanks{Corresponding author: \texttt{mconte@bcamath.org}}}
\author[2]{Christina Surulescu}
\affil[1]{\centerline{\small BCAM - Basque Center for Applied Mathematics} \newline \centerline{\small Alameda de Mazarredo, 14 - E-48009 Bilbao, Spain}}
\affil[2]{\centerline {\small Technische Universit\"at Kaiserslautern, Felix-Klein-Zentrum f\"ur Mathematik} \newline \centerline{\small Paul-Ehrlich-Str. 31 - 67663 Kaiserslautern, Germany}}
\date{\today}                     
\setcounter{Maxaffil}{0}
\renewcommand\Affilfont{\itshape\small}
\maketitle

\begin{abstract}
Starting from kinetic transport equations and subcellular dynamics we deduce a multiscale model for glioma invasion relying on the go-or-grow dichotomy and the influence of vasculature, acidity, and brain tissue anisotropy. Numerical simulations are performed for this model with multiple taxis, in order to assess the solution behavior under several scenarios of taxis and growth for tumor and endothelial cells. An extension of the model to incorporate the macroscopic evolution of normal tissue and necrotic matter allows us to perform tumor grading.
\end{abstract}
 

\section{Introduction}\label{sec:intro}

Glioma is the most frequent type of primary brain tumor. It originates from glia cells and accounts for 78 percent of malignant brain tumors, of which glioblastoma multiforme (GBM) is the most aggressive, being characterized by a fast, infiltrative spread and high proliferation. These features make it very difficult to treat, and go along with a poor survival prognosis \cite{Hayat2011,wrensch}.\\[-2ex]

\noindent
Like in most tumors, glioma development, growth, and invasion are influenced by a multitude of intrinsic and extrinsic factors. Among these, the phenotypic heterogeneity and the biological, physical, and chemical composition of the tumor microenvironment play a decisive role.
Experiments with cultures of glioma cells suggest mutual exclusion of migratory and proliferative behavior, as reviewed e.g. in \cite{Berens1999,Giese2003,Xie2014}; this is known as go-or-grow dichotomy \cite{giese-etal96,Giese1996}. Biological evidence indicates that migratory and proliferative processes share common signaling pathways, suggesting a unique intracellular mechanism that regulates both phenotypes \cite{Giese2003}. Hypoxia is a prominent trait of tumor microenvironment and glioma neoplasms are no exception. It has been suggested (see e.g. \cite{Hoering2012,Xie2014} and references therein) that it is putatively influencing the phenotypic switch between migrating and proliferating behavior - along with other regulating factors, like angiogenesis, ECM production and degradation, etc. Indeed, glioma cells have been observed to move away from highly  hypoxic sites created, for instance, by the occlusion of a capillary and to form so-called pseudopalisade patterns, which are typical for glioblastoma \cite{Brat2004b,Brat2004,Wippold2037}. They have garland-like shapes exhibiting central necrotic zones surrounded by stacks of tumor cells, most of which are actively migrating. As (tumor) cell proliferation is impaired at (too) low pH, this seems to endorse the antagonistic relationship between (transiently) migratory and proliferative phenotypes \cite{Moellering2008}. \\[-2ex]

\noindent	
While cancer cells are able to survive in relatively acidic regions by anaerobic glycolysis, which confers them an advantage against normal cells \cite{VanderHeiden2009,Webb2011}, the large amounts of lactate and alanine they produce during this process can decrease the pH below critical levels. As a consequence,  they initiate (re)vascularization by expressing pro-angiogenic factors, in order to provide adequate supply with blood-transported nutrients \cite{Hardee2012,Wang2019}. They can deter proliferation for migration towards more favorable areas \cite{MartnezZaguiln1996,Webb2011}, and the above mentioned pseudopalisade formation is just one aspect of this complex behavior.\\[-2ex]
   
\noindent
Previous models for glioma invasion have been proposed in (semi)discrete \cite{boettger-etal-12,Hatzikirou2010,Kim2013} or continuous frameworks. Most of the latter are purely macroscopic, describing the evolution of glioma cell density under the influence of surrounding tissue, chemical signals, and/or vasculature, see e.g.  \cite{Hinow,Jbabdi2005,Konukoglu2010,MartnezGonzlez2012,Swanson2011} and the review \cite{Alfonso17}; many of them are versions or extensions of a model proposed by Murray \cite{Murray1989}. More recent continuous models leave behind the classical reaction-diffusion prototype and take into account advection bias of glioma cells in response to environmental cues. Some of these are directly set on the macroscopic scale and rely on balance of mass/flux/momentum \cite{Colombo2015,Hogea2007,kim2009}, others are obtained by more detailed descriptions from lower scale dynamics, as proposed in \cite{PH13} and further developed in \cite{Conte2020,Corbin2018,Engwer2,EHS,Engwer,Hunt2016,Kumar20,Swan2017}. In particular, some of the latter have a multiscale character that is (partially) preserved during the upscaling process from kinetic transport equations (mesoscale) to reaction-diffusion-taxis PDEs (macroscale). As mentioned above, hypoxia is an essential factor in tumor evolution. Early mathematical models for cancer invasion and patterning under acidic conditions were introduced in \cite{Gatenby,McGillen2013,Smallbone2008}. Further PDE-based models, that characterize acid-mediated tumor development, directly related or applicable to pH-influenced glioma spread have been proposed and investigated e.g. in \cite{MartnezGonzlez2012} and \cite{Engwer4,Meral2015-JCSMS,stinner-surulescu-meral}, respectively. Instead, for stochastic multiscale settings see e.g. \cite{Hiremath2015,Athni_Hiremath_2016,AthniHiremath2018} and \cite{HiSu-LNM} for a review also addressing further related models. Tumor heterogeneity has been usually modeled in the continuous setting by describing the dynamics of the correspondingly defined subpopulations of cells, e.g. hypoxic/normoxic/necrotic  \cite{Hinow,MartnezGonzlez2012,Swanson2011} or moving/proliferating \cite{Engwer,Gerlee2012,Hunt2016,Meral2015-JCSMS,Pham2012,stinner-surulescu-uatay,Zhigun2018}. An indirect accounting for a go-grow-recede heterogeneity under the influence of intra- and extracellular acidity was proposed and analyzed in \cite{Athni_Hiremath_2016}; its numerical simulations were able to explain a large variety of invasion patterns.\\[-2ex]

\noindent
The models in \cite{Engwer,Hunt2016} started from a mesoscopic description of glioma density functions for migrating, respectively proliferating cells in interaction with the anisotropic brain tissue, whereby the latter took into account subcellular dynamics of receptor binding to tissue fibers on the microscale. The cells were able to switch between the two phenotypes, the corresponding rates being dependent on space and (in \cite{Hunt2016}) also on the doses of some chemotherapeutic agent administered in order to impair cell motility. In this note we extend those settings by including the dynamics of endothelial cells developing vasculature and of acidity, both being responsible for the phenotypic switch. The upscaling of the kinetic transport PDEs for glioma and endothelial cells leads in the parabolic limit to a system of reaction-advection-diffusion equations featuring several types of taxis, nonlinear myopic diffusion of glioma, and highly complicated couplings between the variables of the model, which has a multiscale character due to (some of) the taxis coefficients encoding information from the lower modeling levels. The subsequent content is organized as follows. Section \ref{model} is concerned with the setup of the model on the subcellular and mesoscopic scales and with the deduction of the macroscopic PDEs. In Section \ref{CoefFun} we concretize the coefficient functions in preparation for the numerical simulations to be performed in Section \ref{simulation}.  In order to facilitate the evaluation of the tumor burden in relation to the necrotic and the normal tissue, Section \ref{sec:extension} includes a model extension accounting for the dynamics of the latter. We conclude with a discussion of the results in Section \ref{conclusion}. The Appendix contains the assessment of the model parameters and a nondimensionalization of the macroscopic PDEs.

\section{Modeling}\label{model}

Relying on the go-or-grow dichotomy, we consider a tumor containing two mutually exclusive subpopulations of glioma cells, which are either migrating or proliferating. The respective states are transient, the tumor can change dynamically its composition, according to the signals received by the cells from their surroundings. From the huge variety of chemical and physical cues present in the extracellular space and influencing the development and spread of cancer \cite{hanahan2011} we focus here on the effects of acidity and cell-tissue interactions. Since acidification and angiogenesis are tightly interrelated and crucial for the tumor evolution, we also model vascularization, by way of endothelial cell dynamics. We develop a multiscale model upon starting from the subcellular level of interactions between cells, acidity, and tissue, setting up the corresponding kinetic transport equations (KTEs) for glioma cells of the two phenotypes and for endothelial cells, and performing a parabolic limit to deduce the macroscopic system of reaction-advection-diffusion PDEs for the involved quantities: total tumor burden (moving + proliferating cells), endothelial cells, and acidity (concentration of protons).

\subsection{Subcellular level}\label{cell level}

On the microscopic scale, we describe the interaction of glioma cells with the extracellular space, more precisely with tissue fibers and protons. Cells exchange information with their environment through various transmembrane entities, e.g. cell surface receptors and ion channels. We will use the former to account for cell-tissue interactions and both for the cell-proton exchange. Indeed, additionally to ion channels and membrane transporters which have been extensively studied in the context of intra- and extracellular pH regulation, there also exist proton-sensing receptors \cite{Holzer2009}, e.g. the G protein-coupled receptors (GPCRs) involved among others in regulating the migration and proliferation of cells in tumor development and wound healing \cite{Justus2013,Weisz2017}. We ignore here all intricate details about the intracellular machinery activated by receptor binding and channel opening and closing. Instead, we see the events of occupying such transmembrane units as triggering the cellular processes leading to migration, proliferation, and phenotypic switch.\\[-2ex]

\noindent
Considering such interactions between cells and soluble as well as unsoluble ligands follows the idea employed in \cite{Kelkel2011,KELKEL2012,Lorenz2014} to build a micro-meso model for tumor invasion with chemo- and haptotaxis and revisited in \cite{Conte2020,Corbin2018,Engwer2,EHS,Engwer,Hunt2016} for cell-tissue interactions, in \cite{Kumar20} for cell-proton interactions leading to pseudopalisade patterns, or in \cite{Engwer4} for a nonlocal micro-macro model with cell-tissue, cell-cell, and cell-protons interactions. The micro-macro framework proposed in \cite{Hiremath2015,Athni_Hiremath_2016,AthniHiremath2018} offers a related, yet different perspective on the interplay between intra- and extracellular acidity and tumor cells.\\[-2ex]

\noindent
We denote by $y_1(t)$ the amount of receptors bound to tissue fibers and by $y_2(t)$ that of transmembrane entities (ion channels, pH-sensing receptors) occupied by protons. The corresponding binding/occupying dynamics is characterized by simple mass action kinetics: 
\begin{equation*}
\begin{split}
\bar{R}_0-(y_1+y_2)\,\,+\,\,\dfrac{Q}{Q^*}\,\,\stackrel[k_1^-]{k_1^+}{\rightleftharpoons}\,\,y_1\\[0.2cm]
\bar{R}_0-(y_1+y_2)\,\,+\,\, \dfrac{S}{S_{c,0}}\,\,\stackrel[k_2^-]{k_2^+}{\rightleftharpoons}\,\,y_2.
\end{split}
\end{equation*}
Here $\bar{R}_0$ is the total amount of receptors on a cell membrane, $Q(x)$ denotes the macroscopic tissue density, depending on the position $x \in \mathbb{R}^n$, $Q^*(x)$ is the reference tissue density (corresponding to its carrying capacity), while $S(t,x)$ is the macroscopic concentration of protons, and $S_{c,0}(x)$ the reference proton concentration. Accordingly, we get the ODE system
\begin{equation*}
\begin{split}
&\dot{y_1}=\dfrac{k_1^+}{Q^*}\,Q(x)\,(\bar{R}_0-(y_1+y_2))-k_1^-y_1\\[0.1cm]
&\dot{y_2}= \dfrac{k_2^+}{S_{c,0}}\,S(t,x)\,(\bar{R}_0-(y_1+y_2))-k_2^-y_2
\end{split}
\end{equation*}
where $k_1^+$ and $k_1^-$ represent the attachment and detachment rates of cells to tissue, respectively, while $k_2^+$ and $k_2^-$ are the corresponding rates in the process of proton binding. As in \cite{Engwer4}, we define $y:=y_1+y_2$ to be the total amount of transmembrane entities occupied by tissue or protons, which allows us to lump together the two ODEs, into
\begin{equation*}
\dot{y}=\left(\dfrac{k_1^+}{Q^*}Q+ \dfrac{k_2^+}{S_{c,0}}S\right)\bar{R}_0-y\left(\dfrac{k_1^+}{Q^*}Q+ \dfrac{k_2^+}{S_{c,0}}S\right)-k_1^-y_1-k_2^-y_2\,.
\end{equation*}
Assuming that $k_1^-=k_2^-=k^-$, we get the microscopic equation for the subcellular dynamics
\begin{equation}
\dot{y}=\left(\dfrac{k_1^+}{Q^*}Q+ \dfrac{k_2^+}{S_{c,0}}S\right)(\bar{R}_0-y)-k^-y\,.
\end{equation}
\noindent Since processes on this scale are much faster than those happening on the macroscopic time scale, they can be assumed to equilibrate rapidly and moreover, that on this scale we can ignore the time dependence of $S$. Rescaling $y/{\bar R_0}\leadsto y$ will further simplify the notation. Then, the unique steady state of the above equation is given by:
\begin{equation*}
y^*=\dfrac{\frac{k_1^+}{Q^*}Q+ \frac{k_2^+}{S_{c,0}}S}{\frac{k_1^+}{Q^*}Q+ \frac{k_2^+}{S_{c,0}}S+k^-}=:\bar{f}\left(Q,S\right)\,.
\end{equation*}
The variable $y$ can be seen as characterizing the 'internal' cellular state. In the kinetic theory of active particles (see e.g. \cite{Bell-KTAP,bellomo2010}) it is called 'activity variable'. In the sequel, we will consider the mesoscopic densities $p(t,x,v,y)$ and $r(t,x,y)$ of migrating and respectively non-moving (thus proliferating) glioma cells, hence both depending on such activity variable $y$. Thereby, $v\in V\subset \R^n$ is the cell velocity vector, with the space $V$ to be closer explained in Subsection \ref{subsec:meso_level} below.\\[-2ex]

\noindent
Further, we assume that the glioma cells follow the tissue gradient, but move away from highly acidic areas. Therefore, we look at the path of a single cell starting at position $x_0$ and moving to position $x$ with velocity $v$ in the (locally) time-invariant density fields $Q$ and $S$, so that $Q(x)=Q(x_0+vt)$, while $S(x)=S(x_0-vt)$. 
Denoting by $z:=y^*-y$ the deviation of $y$ from its steady state, we have:
\begin{equation*}
\dot{z}=
\dfrac{\partial \bar{f}}{\partial Q}\, v\cdot \nabla_x Q-\,\dfrac{\partial \bar{f}}{\partial S}\, v\cdot \nabla_x S-z\left(\dfrac{k_1^+}{Q^*}Q+ \dfrac{k_2^+}{S_{c,0}}S+k^-\right),
\end{equation*}
with
\begin{equation*}
\begin{split}
&\dfrac{\partial \bar{f}}{\partial Q}=\dfrac{1}{Q^*}\dfrac{k_1^+k^-}{\left(\frac{k_1^+}{Q^*}Q+ \frac{k_2^+}{S_{c,0}}S+k^-\right)^2}\\[0.1cm]
&\dfrac{\partial \bar{f}}{\partial S}= \dfrac{1}{S_{c,0}}\dfrac{k_2^+k^-}{\left(\frac{k_1^+}{Q^*}Q+ \frac{k_2^+}{S_{c,0}}S+k^-\right)^2}\, ,
\end{split}
\end{equation*}
thus the equation for $z$ is given by:
\begin{equation}
\dot{z}=-z\left(\dfrac{k_1^+}{Q^*}Q+ \dfrac{k_2^+}{S_{c,0}}S+k^-\right)+\dfrac{k^-}{\left(\frac{k_1^+}{Q^*}Q+ \frac{k_2^+}{S_{c,0}}S+k^-\right)^2}\, \left(\dfrac{k_1^+}{Q^*}v\cdot \nabla_x Q- \dfrac{k_2^+}{S_{c,0}}v\cdot \nabla_x S\right)=:G(z,Q,S)\,.
\end{equation}
To simplify the writing, we denote 
$B(Q,S):=\left(\dfrac{k_1^+}{Q^*}Q+ \dfrac{k_2^+}{S_{c,0}}S+k^-\right)$.\\

\subsection{Mesoscopic level}\label{subsec:meso_level}

We model the mesoscale behavior of glioma and endothelial cells with the aid of kinetic transport equations (KTEs) describing velocity-jump processes and taking into account the subcellular dynamics. Concretely, we consider the following cell density functions:
\begin{itemize}
	\item $p(t,x,v,y)$ for moving glioma cells;
	\item $r(t,x,y)$ for non-moving, hence (in virtue of the go-or-grow dichotomy) proliferating glioma cells;
	\item $w(t,x,\vartheta)$ for endothelial cells (ECs) forming capillaries, 
\end{itemize}
with the time and space variables $t>0$ and $x\in\R^n$, velocities $v\in V=s\mathbb{S}^{n-1}$ and $\vartheta\in \Theta :=\sigma \mathbb{S}^{n-1}$, and activity variable $y\in Y=(0,1)$. These choices mean that we assume for glioma and ECs constant speeds $s>0$ and $\sigma >0$, respectively, where $\mathbb{S}^{n-1}$ denotes the unit sphere in $\mathbb{R}^n$. As in \cite{Engwer2,EHS,Engwer,Hunt2016}, in the sequel we will work with the deviation $z=y^*-y\in Z\subseteq (y^*-1,y^*)$ rather then with $y$.\\[-2ex]

\noindent
The corresponding macroscopic cell densities are denoted by $M(t,x)$, $R(t,x)$ and $W(t,x)$, respectively, and we use the notation $N(t,x):=M(t,x)+R(t,x)$ for the space-time varying macroscopic total tumor burden.\\[-2ex]

\noindent
The kinetic transport equation for the motile glioma phenotype is given by:\\
\begin{equation}
\partial_t p +\nabla_x\cdot\,(v p) +\partial_z(G(z,Q,S)p)=\mathcal{L}_p[\lambda(z)]p+\beta(S)\dfrac{q}{\omega}r-\alpha (w,S)p-l_m(N)p\,,
\label{peq}
\end{equation}
where $\mathcal{L}_p[\lambda(z)]p$ denotes the turning operator, describing velocity changes. In particular, such changes are due to contact guidance: the cells have a tendency to align to the brain tissue anisotropy, mainly associated with white matter tracts. As in previous models of this type, $\mathcal{L}_p[\lambda]p$ is a Boltzmann-like integral operator of the form
 \begin{equation} 
\mathcal{L}_p[\lambda(z)]p=-\lambda(z)p+\lambda(z)\int_V K(x,v)\,p(v')\,dv'\,,
 \label{turn_rate}
 \end{equation}
where $\lambda(z):=\lambda_0-\lambda_1z \ge 0$ is the cell turning rate  depending on the microscopic variable $z$, while $\lambda_0$ and $\lambda_1$ are positive constants. The integral term describes the reorientation of cells from any previous velocity $v'$ to a new velocity $v$ upon interacting with the tissue fibers. This is controlled by the turning kernel $K(x,v)$, which we assume to be independent on the incoming velocity $v'$. In particular, we consider that the dominating directional cue is given by the orientation of tissue fibers and take as e.g. in \cite{Engwer2,PH13}
$K(x,v):=\frac{q(x,\hat v)}{\omega}$, where $\hat v=\frac{v}{|v|}$ and $q(x,\theta)$ with $\theta \in \mathbb{S}^{n-1}$ is the orientational distribution of the fibers, normalized by 	$\omega =s^{n-1}$. It encodes the individual brain structure obtained by diffusion tensor imaging (DTI); concrete choices will be provided in Section \ref{CoefFun}. We assume the tissue to be undirected, hence $q(x,\theta )=q(x,-\theta)$ for all $x\in \R^n$. For later reference, we introduce the notations
\begin{align*}
&\mathbb E_q(x):=\int _{\Sbb} \theta q(x,\theta ) d\theta \\
&\mathbb V_q(x):=\int _{\Sbb} (\theta -\mathbb E_q)\otimes (\theta -\mathbb E_q)\ q(x,\theta ) d\theta
\end{align*}
for the mean fiber orientation and the variance-covariance matrix for the orientation distribution of tissue fibers, respectively. Notice that the above symmetry of $q$ implies $\mathbb E_q=0$.
\\[-2ex]

\noindent
The term $\beta(S)\dfrac{q}{\omega}r$ in \eqref{peq} describes the phenotypic switch $r \to p$;  the rate $\beta$ depends on $S$, since in a too acidic environment the cells are supposed to stop proliferation and migrate towards regions with higher pH (see Section \ref{sec:intro}). Therefore, we assume that $\beta(S)$ is an increasing function of the proton concentration $S$ and we require $\beta (S)>0$, since there will always be some proliferating cells switching into a migratory regime \footnote{otherwise the tumor would stay confined, which is not the case for glioblastoma}.\\[-2ex]

\noindent
 The terms $\alpha (w,S)p$ and $l_m(N)p$ model the phenotypic switch $p \to r$ due to environmental signals. Thus, the former term depends on the proton concentration $S$ and on the mesoscopic EC density $w$ and describes the adoption of a proliferative phenotype when there are enough oxygen and nutrient supply, while the acidity remains below a certain threshold. The latter term models the switch to proliferation caused by the glioma cell population being too crowded to allow effective migration (but still allowing some limited proliferation). In order not to complicate too much the model we do not explicitly account for cell recession or quiescence.\\[-2ex]

\noindent
The evolution of proliferating tumor cells is characterized by the integro-differential equation
\begin{equation}
\partial_t r=(\alpha (w,S)+l_m(N))\int_{V}p(t,x,v,z)\,dv+\mu(W,N,S)\int_Z\chi(x,z,z') \dfrac{Q(x)}{Q^*}r(t,x,z')dz'-(\beta(S)+\gamma(S))r\,,
\label{req}
\end{equation}
where, additionally to the already described switch terms, we model intrinsic proliferation and death. $\mu(W,N,S)$\\ $\int_Z\chi(x,z,z') \frac{Q(x)}{Q^*(x)}r(z')dz'$ describes as in \cite{EHS,Hunt2016} proliferation triggered by cell receptor binding to tissue. The proliferation rate $\mu(W,N,S)$ depends on the total macroscopic  tumor density $N$, on the concentration of protons $S$, and on the vasculature. In the integral operator, the kernel $\chi (x,z,z')$ characterizes the transition from state $z'$ to state $z$ during such proliferation-initiating interaction at position $x$. No further conditions are required on $\chi$, we only assume that the nonlinear proliferative operator is uniformly bounded in the $L^2$-norm, which is reasonable, in view of space-limited cell division. The last term $-\gamma(S)r$ describes acid-induced death of glioma cells when the pH value drops below a certain threshold.\\[-2ex]

\noindent
The KTE for ECs is given by:\\
\begin{equation}
\partial_t w+\nabla_x\cdot\,(\vartheta w)=\mathcal{L}_w[\eta]w+\mu_W(W,Q)w\,,
\label{weq}
\end{equation}
where the turning operator $\mathcal{L}_w[\eta]w$ describes changes in the orientation of ECs. It is well-established (see e.g. \cite{hanahan2011}) that tumor cells produce angiogenic signals acting as chemoattractants for ECs. Less known is whether such signals are expressed by proliferating rather than moving cells or by both phenotypes. There is, however, evidence that hypoxia induces  production of VEGF and other angiogenic cytokines (see e.g. \cite{Chouaib2012,Semenza2009,Xu2001} and references therein). Since cancer cells are highly glycolytic (which leads to acidification), and increased glucose metabolism is selected for in proliferating cells \cite{lunt}, we assume that pro-angiogenic signals are mainly produced by proliferating cells. With the aim of avoiding the introduction of a new variable for the concentration of such chemoattractants, we let the ECs be attracted by the proliferating glioma cells as main sources therewith. This translates into the following form of the turning operator acting on the right hand side in \eqref{weq}: 
\begin{equation*}
\mathcal{L}_w[\eta]w=-\eta(x,\vartheta,R)w(t,x,\vartheta)+\int_{\Theta }\dfrac{1}{|\Theta |}\eta(x,\vartheta',R)w(\vartheta')d\vartheta'\,,
\end{equation*}
where for the turning kernel modeling ECs reorientations we simply took a uniform density function over the unit sphere $\mathbb{S}^{n-1}$ and 
\begin{equation}
\eta(x,\vartheta,R)=\eta_0(x)e^{-a(R)D_tR}
\label{W_turning_rate}
\end{equation} 
represents the turning rate of ECs. It depends on the macroscopic density of proliferating tumor cells $R$ and on the pathwise gradient $D_tR=\partial _tR+\vartheta \cdot \nabla R$. The coefficient function $a(R)$ is related to the interactions between ECs and proliferating glioma, more precisely ECs and pro-angiogenic signals produced by the latter. This can be described for instance via equilibrium of EC receptor binding. This way to include directional bias provides an alternative to that using a transport term with respect to the activity variable $z$ in \eqref{peq} and it has been introduced in \cite{Othmer2002} in the context of bacteria movement and recently used in \cite{Kumar20} for a model for glioma pseudopalisade patterning. Under certain conditions, the relation between the two approaches was established rigorously for bacteria chemotaxis in \cite{Perthame2016} and investigated more formally for glioma repellent pH-taxis in \cite{Kumar20}. \\[-2ex]

\noindent
Finally, $\mu_W(W,Q)w$ is the proliferation term for ECs. Besides the total population of ECs irrespective of their orientation, $\mu_W$ is supposed to depend on the available macroscopic tissue $Q$. More details about the concrete choices of the coefficient functions involved in \eqref{peq}, \eqref{req}, and \eqref{weq} will be provided in Section \ref{CoefFun}.

\subsection{Parabolic scaling of the mesoscopic model}\label{subsec:upscaling}
\label{scaling}
\noindent 
The high dimensionality of the KTE system (\ref{peq}), (\ref{req}), and (\ref{weq}) makes its numerical simulations too expensive. Moreover, clinicians are interested in the macroscopic evolution of the tumor along with its vascularization and acidity profile. Therefore, it is convenient to deduce effective equations for the macroscopic dynamics of ECs and total tumor burden $N=R+M$. The PDE for the proton concentration $S$ is already macroscopic and does not need to be upscaled. We rescale the time and space variables as $t\to\eps^2t$ and $x\to\eps x$. Moreover, the proliferation terms in (\ref{req}) and \eqref{weq} and the death term in (\ref{req}) will be scaled by $\eps^2$ in order to account for the mitotic and apoptotic events taking place on a much larger time scale than migration and switching from moving to non-moving regimes. Hence, (\ref{peq}), (\ref{req}), and (\ref{weq}) become: \\[-2ex]
\begin{align}
&\eps^2 \partial _tp+\eps \nabla _x\cdot (vp)-\partial_z\Bigg (\Big (zB(Q,S)-\frac{\eps  k^-}{B(Q,S)^2}\left(\dfrac{k_1^+}{Q^*}v\cdot \nabla Q- \dfrac{k_2^+}{S_{c,0}}v\cdot \nabla S\right)p\Bigg )=\mathcal{L}_p[\lambda(z)]p+\beta(S)\dfrac{q}{\omega}r\notag \\
&\hspace{12cm} -\alpha (w,S)p-l_m(N)p\label{peq-eps}\\
&\eps^2\partial_t r=(\alpha (w,S)+l_m(N))\int_{V}p(v)\,dv+\eps ^2\mu(W,N,S)\int_Z\chi(x,z,z') \dfrac{Q(x)}{Q^*}r(t,x,z')dz'-(\beta(S)+\eps ^2\gamma(S))r\,,
\label{req-eps}\\
&\eps^2\partial_t w+\eps \nabla_x\cdot\,(\vartheta w)=\mathcal{L}_w[\eta^\eps ]w+\eps ^2\mu_W(W,Q)w\,,
\label{weq-eps}
\end{align}	
with 
\begin{align}
&\eta^{\eps}(x,\vartheta ,R)=\eta _0(x)\exp \Big (-a(R)(\eps ^2\partial _tR+\eps \vartheta \cdot \nabla R)\Big ).
\end{align}
\noindent
As in \cite{Engwer2,EHS,Engwer}, we define the following moments:
\begin{equation*}
\begin{split}
&m(t,x,v)=\int_Zp(t,x,v,z)dz,\quad\quad R(t,x)=\int_Zr(t,x,z)dz,\quad \quad
m^z(t,x,v)=\int_Zzp(t,x,v,z)dz,\\[0.2cm]
&R^z(t,x)=\int_Zzr(t,x,z)dz,\quad\quad  M(t,x)=\int_Vm(t,x,v)dv,  \quad\quad W(t,x)=\int_{\Theta } w(t,x,\vartheta)d\vartheta,\\[0.2cm] 
&M^z(t,x)=\int_Vm^z(t,x,v)dv\,
\end{split}
\end{equation*}
and neglect the higher order moments w.r.t. the variable $z$, in virtue of the subcellular dynamics being much faster than the events on the higher scales, hence $z \ll 1$. We assume the functions $p$ and $r$ to be compactly supported in the phase space $\mathbb{R}^n \times V \times Z$ and $w$ to be compactly supported in $\mathbb{R}^n \times \Theta $.\\[-2ex]

\noindent
We first integrate equation (\ref{peq-eps}) with respect to $z$, getting the following equation for $m(t,x,v)$:
\begin{equation}
\eps^2\partial_t m+\eps \nabla_x\cdot(v m)=-\lambda_0\left(m-\dfrac{q}{\omega}M\right)+\lambda_1\left(m^z-\dfrac{q}{\omega}M^z\right)+\beta(S)\dfrac{q}{\omega}R-\alpha (w,S)m-l_m(N)m\,.
\label{meq}
\end{equation}

\noindent Integrating equation (\ref{req-eps}) with respect to $z$ we get
\begin{equation*}
\eps^2\partial_t R(t,x)=\left(\alpha (w,S)+l_m(N)\right)M(t,x)+\eps^2\int_Z\mu(W,N,S)\int_Z\chi(z,z',x)r(z')\dfrac{Q(x)}{Q^*}dz'dz-\left(\beta(S)+\eps^2\gamma(S)\right)R(t,x)\,.\end{equation*}
Using the fact that $\chi(z,z',x)$ is a probability kernel with respect to $z$, the previous equation for $R(t,x)$ reduces to:
\begin{equation}
\eps^2\partial_t R=\left(\alpha (w,S)+l_m(N)\right)M+\eps^2\mu(W,N,S) \dfrac{Q(x)}{Q^*}\,R-\left(\beta(S)+\eps ^2\gamma(S)\right)R\,.
\label{Req}
\end{equation}

\noindent Then, we multiply equation (\ref{peq-eps}) by $z$ and integrate it w.r.t. $z$, obtaining

\begin{equation*}
\begin{split}
\eps ^2\partial_t m^z&+\eps \nabla_x\cdot(v m^z)-\int_Zz\partial_z\left[\left(zB(Q,S)-\dfrac{\eps k^-}{B(Q,S)^2}\left(\dfrac{k_1^+}{Q^*}v\cdot \nabla Q- \dfrac{k_2^+}{S_{c,0}}v\cdot \nabla S\right)\right)p(z)\right]dz\\[0.2cm]&=\int_Zz\mathcal{L}_p[\lambda(z)]p(z)dz+\beta(S)\dfrac{q}{\omega}R^z-(\alpha (w,S)+l_m(N))m^z\,.
\end{split}
\end{equation*}
The calculation of the integral term on the left hand side leads to the following equation for $m^z(t,x,v)$:
\begin{align}
\eps ^2\partial_t m^z+\eps \nabla_x\cdot(v m^z)+B(Q,S)m^z-\dfrac{\eps k^-}{B(Q,S)^2}\left(\dfrac{k_1^+}{Q^*}v\cdot \nabla Q- \dfrac{k_2^+}{S_{c,0}}v\cdot \nabla S\right)m&=-\lambda_0m^z+\lambda_0\dfrac{q}{\omega}M^z+\beta(S)\dfrac{q}{\omega}R^z\notag \\
&-(\alpha (w,S)+l_m(N))m^z\,.
\label{mzeq}
\end{align}
\noindent Applying the same procedure to equation (\ref{req-eps}) we obtain an equation for $R^z(t,x)$: 
\begin{equation}
\eps ^2\partial_t R^z=\left(\alpha (w,S)+l_m(N)\right)M^z+\eps ^2\mu(W,N,S) \dfrac{Q(x)}{Q^*}\int _Z\int _Zz\chi (x,z,z')r(z')dz'dz-\left(\beta(S)+\eps ^2\gamma(S)\right)R^z\,.
\label{Rzeq}
\end{equation}

\noindent
We consider Hilbert expansions for the previously introduced moments:
\begin{equation*}
\begin{split}
&m(t,x,v)=\sum_{k=0}^{\infty}\eps^km_k,\quad\quad  R(t,x)=\sum_{k=0}^{\infty}\eps^kR_k,\quad\quad 
m^z(t,x,v)=\sum_{k=0}^{\infty}\eps^km^z_k\quad \quad 
R^z(t,x)=\sum_{k=0}^{\infty}\eps^kR^z_k, \\[0.2cm] 
&M(t,x)=\sum_{k=0}^{\infty}\eps^kM_k,\quad \quad w(x,t,\vartheta)=\sum_{k=0}^{\infty}\eps^kw_k,\quad\quad
M^z(t,x)=\sum_{k=0}^{\infty}\eps^kM^z_k,\quad\quad  W(x,t)=\sum_{k=0}^{\infty}\eps^kW_k\,.
\end{split}
\end{equation*}
For the subsequent calculations it will be useful to Taylor-expand the coefficient functions involving any of $w$, $W$, $R$, $M$ or $N$ in the scaled equations \eqref{meq}-\eqref{Rzeq} and \eqref{weq-eps}:
\begin{align*}
&\alpha (w,S)=\alpha (w_0,S)+\partial_w\alpha (w_0,S)\,(w-w_0)+\dfrac{1}{2}\partial^2_{ww}\alpha (w_0,S)\,(w-w_0)^2+O(|w-w_0|^3),\\[0.5ex]
&l_m(N)=l_m(N_0)+l_m'(N_0)\,(N-N_0)+\dfrac{1}{2}l''_m(N_0)\,(N-N_0)^2+O(|N-N_0|^3),\\[0.5ex]
&\mu (W,N,S)=\mu (W_0,N_0,S)+\partial _W\mu (W_0,N_0,S)(W-W_0)+\partial _N\mu (W_0,N_0,S)(N-N_0)+O(\eps ^2),\\[0.5ex]
&\mu_W(W,Q)=\mu_W(W_0,Q)+\partial_W\mu_W(W_0,Q)(W-W_0)+O(|W-W_0| ^2),\\[0.5ex] 
&\eta ^\eps (x,\vartheta, R)=\eta _0(x)\Big [1-\eps a(R)\vartheta \cdot \nabla R+\eps ^2\Big (-a(R)\partial_tR+\frac{1}{2}(a(R))^2(\vartheta \cdot \nabla R)^2\Big )+O(\eps^3)\Big ],\\[0.5ex]
&a(R)=a(R_0)+a'(R_0)(R-R_0)+\frac{1}{2}a''(R_0)(R-R_0)^2+O(|R-R_0|^3).
\end{align*}

\noindent
Then, equating the powers of $\eps $ in the scaled equations \eqref{meq}-\eqref{Rzeq} and \eqref{weq-eps}, we obtain:\\[-2ex]

\noindent $\eps^0$ terms:
\begin{align}
&0=-\lambda_0\left(m_0-\dfrac{q}{\omega}M_0\right)+\lambda_1\left(m_0^z-\dfrac{q}{\omega}M_0^z\right)+\beta(S)\dfrac{q}{\omega}R_0-\Big (\alpha (w_0,S)+l_m(N_0)\Big )m_0, \label{m0eq}\\
&0=\Big (\alpha (w_0,S)+l_m(N_0)\Big )M_0-\beta(S)R_0, \label{R0eq}\\[0.5ex]
&0=-\Big(B(Q,S)+\lambda_0+\alpha (w_0,S)+l_m(N_0)\Big )m_0^z+\lambda _0\dfrac{q}{\omega}M_0^z+\beta(S)\dfrac{q}{\omega}R_0^z, \label{m0zeq}\\[0.5ex]
&0=\Big (\alpha (w_0,S)+l_m(N_0)\Big )M_0^z-\beta(S)R_0^z, \label{R0zeq}\\[0.5ex]
&0=\eta_0(x) (S_n^\sigma W_0-w_0), \label{w0eq}
\end{align}
where $S_n^\sigma:=\frac{1}{|\Theta |}=\frac{\sigma ^{1-n}}{|\mathbb S^{n-1}|}$.\\[-2ex]

\noindent $\eps^1$ terms:
\begin{align}
&\nabla_x\cdot(v m_0)=-\lambda_0\left(m_1-\dfrac{q}{\omega}M_1\right)+\lambda_1\left(m_1^z-\dfrac{q}{\omega}M_1^z\right)+\beta(S)\dfrac{q}{\omega}R_1-\Big (\alpha (w_0,S)+l_m(N_0)\Big )m_1\nonumber \\
&\hspace{2cm}-\Big (\partial_w\alpha (w_0,S)w_1+l'_m(N_0)N_1\Big )m_0, \label{m1eq}\\[0.5ex]
&0=\Big (\alpha (w_0,S)+l_m(N_0)\Big )M_1+\Big (\partial_w \alpha (w_0,S)w_1+l'_m(N_0)N_1\Big )M_0-\beta(S)R_1, \label{R1eq}\\[0.5ex]
&\nabla_x\cdot(v m_0^z)=-B(Q,S)m_1^z+\dfrac{k^-}{B(Q,S)^2}\left(\dfrac{k_1^+}{Q^*}v\cdot \nabla Q- \dfrac{k_2^+}{S_{c,0}}v\cdot \nabla S\right)m_0-\lambda_0\Big (m_1^z-\dfrac{q}{\omega}M_1^z\Big )+\nonumber \\[0.5ex] 
&\hspace{2cm}+\beta(S)\dfrac{q}{\omega}R_1^z-\Big (\alpha (w_0,S)+l_m(N_0)\Big )m_1^z-\Big (\partial_w\alpha (w_0,S)w_1+l'_m(N_0)N_1\Big )m_0^z, \label{m1zeq}\\[0.5ex]
&0=\Big (\alpha (w_0,S)+l_m(N_0)\Big )M_1^z+\Big (\partial_w\alpha (w_0,S)w_1+l'_m(N_0)N_1\Big )M_0^z-\beta(S)R_1^z, \label{R1zeq}\\[0.5ex]
&\nabla_x \cdot(\vartheta w_0)=\eta_0(x)\Big (S_n^\sigma W_1-w_1\Big ) +\eta_0(x)a(R_0)\vartheta \cdot\nabla R_0 \,w_0 -S_n^\sigma\eta_0(x)a(R_0)\int_{\Theta }w_0(\vartheta')\vartheta' d\vartheta'\cdot \nabla R_0. \label{w1eq}
\end{align}

\noindent $\eps^2$ terms:
\begin{align}
&\partial_tm_0+\nabla_x\cdot(v m_1)=-\lambda_0\left(m_2-\dfrac{q}{\omega}M_2\right)+\lambda_1\left(m_2^z-\dfrac{q}{\omega}M_2^z\right)+\beta(S)\dfrac{q}{\omega}R_2-\Big (\alpha (w_0,S)+l_m(N_0)\Big )m_2\nonumber \\[0.5ex] 
&\hspace{3.3cm}-\Big (\partial_w\alpha (w_0,S)w_1+l'_m(N_0)N_1\Big )m_1-\Big (\partial_w\alpha (w_0,S)w_2+l'_m(N_0)N_2\Big )m_0\nonumber \\[0.5ex]
&\hspace{3.3cm}-\dfrac{1}{2}\Big (\partial^2_{ww}\alpha (w_0,S)w_1^2+l''_m(N_0)N_1^2\Big )m_0, \label{m2eq}\\[0.5cm]
&\partial_t R_0=\Big (\alpha (w_0,S)+l_m(N_0)\Big )M_2+\Big (\partial_w \alpha (w_0,S)w_1+l'_m(N_0)N_1\Big )M_1\nonumber \\[0.5ex]
&\hspace{1cm}+\dfrac{1}{2}\Big (\partial^2_{ww} \alpha (w_0,S)w_1^2+l''_m(N_0)N_1^2\Big )M_0
+\Big (\partial_w \alpha (w_0,S)w_2+l'_m(N_0)N_2\Big )M_0\nonumber \\[0.6ex]
&\hspace{1cm}+\mu(W_0,N_0,S) \dfrac{Q(x)}{Q^*}R_0-\beta(S)R_2-\gamma(S)R_0, \label{R2eq}
\end{align}
\begin{align}
&\partial_t w_0+\nabla_x\cdot(\vartheta w_1)=
\eta_0(x)\Big [\Big (a(R_0)\partial_tR_0+a(R_0)\vartheta\cdot\nabla R_1 +a'(R_0)R_1\vartheta\cdot\nabla R_0-\dfrac{1}{2}(a(R_0)\vartheta\cdot\nabla R_0)^2\Big )\,w_0\notag \\[0.3cm]
&\hspace{3.1cm}+a(R_0)\vartheta\cdot\nabla R_0 \,w_1-w_2\Big ]\nonumber \\[0.3cm]
&\hspace{3.1cm} +S_n^\sigma\eta_0(x)\Big [W_2-a(R_0)\int_{\Theta }\vartheta'w_1(\vartheta') d\vartheta'\cdot\nabla R_0 -a(R_0)\partial_tR_0W_0\nonumber \\[0.3cm]
&\hspace{3.1cm}-\int_{\Theta }\vartheta'w_0(\vartheta')d\vartheta'\cdot \Big(a(R_0)\nabla R_1 +a'(R_0)R_1\nabla R_0 \Big )\nonumber \\[0.3cm]
&\hspace{3.1cm}+\dfrac{1}{2}\int_{\Theta }(a(R_0)\vartheta'\cdot\nabla R_0)^2w_0(\vartheta')d\vartheta'\Big ]+\mu_W(W_0,Q)w_0\,\nonumber \\[0.3cm]
&\hspace{3.1cm}=\eta_0(x)\Big [\Big ( a(R_0)\partial_tR_0+a(R_0)\vartheta\cdot\nabla R_1 +a'(R_0)R_1\vartheta\cdot\nabla R_0-\dfrac{1}{2}(a(R_0)\vartheta\cdot\nabla R_0)^2\Big )\,w_0\notag \\[0.3cm]
&\hspace{3.1cm}+a(R_0)\vartheta\cdot\nabla R_0 \,w_1-w_2\Big ]\nonumber \\[0.3cm]
&\hspace{3.1cm} +S_n^\sigma\eta_0(x)\Big [W_2-a(R_0)\int_{\Theta }\vartheta'w_1(\vartheta') d\vartheta'\cdot\nabla R_0-a(R_0)\partial_tR_0W_0\nonumber \\[0.3cm]
&\hspace{3.1cm}+\dfrac{1}{2}\int_{\Theta }(a(R_0)\vartheta'\cdot\nabla R_0)^2w_0(\vartheta')d\vartheta'\Big ]+\mu_W(W_0,Q)w_0\,,\label{w2eq}
\end{align}
due to \eqref{w0eq}.\\[-2ex]

\noindent From (\ref{R0eq}) we get
\begin{equation}
R_0(t,x)=\dfrac{\alpha (w_0,S)+l_m(N_0)}{\beta(S)}M_0(t,x)
\label{R0eq_1}
\end{equation}
and from (\ref{R0zeq}) we get
\begin{equation}
R_0^z(t,x)=\dfrac{\alpha (w_0,S)+l_m(N_0)}{\beta(S)}M_0^z(t,x)\,.
\label{R0zeq_1}
\end{equation}
\noindent Integrating (\ref{m0zeq}) with respect to $v$, we have

\begin{equation*}
0=-B(Q,S)M_0^z+\beta(S)R_0^z-\alpha (w_0,S)M_0^z-l_m(N_0)M_0^z,
\end{equation*}
from which by using (\ref{R0zeq_1}) we obtain 

\begin{align}
&M_0^z=0 \label{m0zeq_1}\\[0.2cm]
&R_0^z=0 \label{R0zeq_2}\,.
\end{align}
These, together with (\ref{m0zeq}), lead to 
\begin{equation}
m_0^z=0\,.
\label{m0zeq_2}
\end{equation}
From \eqref{m0eq} and \eqref{m0zeq} we obtain:

\begin{equation*}
\begin{split}
&0=-\lambda_0m_0+\lambda_0\dfrac{q}{\omega}M_0+\Big (\alpha (w_0,S)+l_m(N_0)\Big )\dfrac{q}{\omega}M_0-\Big (\alpha (w_0,S)-l_m(N_0)\Big )m_0\\
&\Rightarrow\quad 0=\left(\dfrac{q}{\omega}M_0-m_0\right)\Big (\lambda_0+\alpha (w_0,S)+l_m(N_0)\Big )\,.
\end{split}
\end{equation*}
Since $(\lambda_0+\alpha (w_0,S)+l_m(N_0))\ne 0$ for any $N_0, S, w_0$, we obtain
\begin{equation}
m_0=\dfrac{q}{\omega}M_0\,.
\label{m0eq_1}
\end{equation}
From equation (\ref{w0eq}) we see that
\begin{equation}
w_0=S_n^\sigma W_0,
\label{w0eq_1}
\end{equation}
thus $w_0$ depends on the (constant) speed $\sigma$, but not on the direction $\theta\in \mathbb S^{n-1}$.\\

\noindent Now, turning to the equations stemming from the $\eps^1$-terms, from (\ref{R1zeq}) we get:
\begin{equation}
R_1^z(x,t)=\dfrac{\alpha (w_0,S)+l_m(N_0)}{\beta(S)}M_1^z(x,t)\,.
\label{R1zeq_1}
\end{equation}
Integrating (\ref{m1zeq}) with respect to $v$, we have upon using \eqref{m0eq_1} and \eqref{m0zeq_2}:

\begin{equation*}
\begin{split}
0=&-B(Q,S)M_1^z+\dfrac{k^-}{B(Q,S)^2}M_0 s^n\mathbb E_q\cdot \left(\dfrac{k_1^+}{Q^*}\nabla Q- \dfrac{k_2^+}{S_{c,0}}\nabla S\right)+ \beta(S)R_1^z-\Big (\alpha (w_0,S)+l_m(N_0)\Big )M_1^z.
\end{split}
\end{equation*}
Using the fact that the tissue is undirected, i.e. $\mathbb E_q=0$ and  \eqref{R1zeq_1}, the previous equation leads to:
\begin{equation}
M_1^z=0\,
\label{m1zeq_1}
\end{equation}
and whence
\begin{equation}
R_1^z=0\,.
\label{R1zeq_2}
\end{equation}
Now plugging these results into \eqref{m1zeq} we get

\begin{equation*}
\begin{split}
&0=-B(Q,S)m_1^z+\dfrac{k^-}{B(Q,S)^2}\left(\dfrac{k_1^+}{Q^*}v\cdot \nabla Q- \dfrac{k_2^+}{S_{c,0}}v\cdot \nabla S\right)m_0-\lambda_0m_1^z-\Big (\alpha (w_0,S)+l_m(N_0)\Big )m_1^z\\[0.3cm]
&\Rightarrow \quad m_1^z \left[B(Q,S)+\lambda_0+\alpha (w_0,S)+l_m(N_0)\right]=\dfrac{k^-}{B(Q,S)^2}\left(\dfrac{k_1^+}{Q^*}v\cdot \nabla Q- \dfrac{k_2^+}{S_{c,0}}v\cdot \nabla S\right)m_0\,.
\end{split}
\end{equation*}
We denote 
\begin{equation}\label{grosses-F}
F(Q,S):=\dfrac{k^-}{B(S,Q)^2\left[B(Q,S)+\lambda_0+\alpha (w_0,S)+l_m(N_0)\right]}
\end{equation}
and obtain therewith the following expression for $m_1^z$:
\begin{equation}
m_1^z=F(Q,S)\left(\dfrac{k_1^+}{Q^*}v\cdot \nabla Q- \dfrac{k_2^+}{S_{c,0}}v\cdot \nabla S\right)m_0.
\label{m1zeq_2}
\end{equation}

\noindent
From equation (\ref{R1eq}) we have
\begin{equation}
R_1=\dfrac{\alpha (w_0,S)+l_m(N_0)}{\beta(S)}M_1+\dfrac{\partial_w\alpha (w_0,S)w_1+l'_m(N_0)N_1}{\beta(S)}M_0.
\label{R1eq_1}
\end{equation}
Using (\ref{m1eq}) we derive
\begin{equation*}
\begin{split}
\nabla_x\cdot(v m_0)=&-\lambda_0\left(m_1-\dfrac{q}{\omega}M_1\right)+\lambda_1m_1^z+\Big (\alpha (w_0,S)+l_m(N_0)\Big )\dfrac{q}{\omega}M_1+\Big (\partial_w\alpha (w_0,S)w_1+l'_m(N_0)N_1\Big )\dfrac{q}{\omega}M_0\\[0.3cm]
&-\Big (\alpha (w_0,S)+l_m(N_0)\Big )m_1-\Big (\partial_w\alpha (w_0,S)w_1+l'_m(N_0)N_1\Big )m_0, 
\end{split}
\end{equation*}
thus by \eqref{m0eq_1}
\begin{align}
\bar{\mathcal{L}}_m[\lambda_0+\alpha (w_0,S)+l_m(N_0)]m_1&:=-\Big (\lambda_0+\alpha (w_0,S)+l_m(N_0)\Big )m_1+\Big (\lambda_0+\alpha (w_0,S)+l_m(N_0)\Big )\dfrac{q}{\omega}M_1\nonumber\\[0.3cm]
&=\nabla_x\cdot(v m_0)-\lambda_1m_1^z\,.\label{pseudo-L-first}
\end{align}
In order to get an explicit expression for $m_1$, we would like to invert the operator $\bar{\mathcal{L}}_m[\lambda_0+\alpha (w_0,S)+l_m(N_0)]$. As e.g. in \cite{PH13,Engwer2}, we define it on the weighted $L^2$-space $L^2_q(V)$, in which the measure $dv$ is weighted by $q(x,\hat v)/\omega$. In particular, $L^2_q(V)$ can be decomposed as $L^2_q(V)=<q/\omega>\oplus<q/\omega>^\perp$. Due to the properties of the chosen turning kernel, $\bar{\mathcal{L}}_m[\lambda_0+\alpha (w_0,S)+l_m(N_0)]$ is a compact Hilbert-Schmidt operator with kernel $<q/\omega>$. We can therefore calculate its pseudo-inverse on $<q/\omega>^\perp$. \\[-2ex]

\noindent
Thus, to determine $m_1$ from \eqref{pseudo-L-first} we need to check the solvability condition
\begin{equation*}
\int_V \left[\nabla_x\cdot(v m_0)-\lambda_1m_1^z \right]dv=0\,.
\end{equation*}
This holds thanks to the above results and to the symmetry of $q(x,\hat v)$. Therefore, we obtain from \eqref{pseudo-L-first} and \eqref{m1zeq_2}
\begin{equation}
m_1=-\dfrac{1}{\lambda_0+\alpha (w_0,S)+l_m(N_0)}\left[\nabla_x\cdot(v m_0)-\lambda_1F(Q,S)\left(\dfrac{k_1^+}{Q^*}v\cdot \nabla Q- \dfrac{k_2^+}{S_{c,0}}v\cdot \nabla S\right)m_0\right]
\end{equation}
and
\begin{equation}
M_1=0\,.
\label{m1eq_1}
\end{equation}
On the other hand, \eqref{w1eq} and \eqref{w0eq_1} give:
\begin{equation}\label{eq:eqq}
\nabla_x \cdot (\vartheta w_0)=-\eta_0(x)w_1+\eta_0(x)a(R_0)\vartheta \cdot \nabla R_0 w_0+S_n^\sigma\eta_0(x)W_1\,.
\end{equation}
Likewise, we observe that the operator $\bar{\mathcal{L}}_w[\eta_0]w_1:=-\eta_0(x)w_1+S_n^\sigma \eta_0(x)W_1$ can be inverted, so that \eqref{eq:eqq} leads to
\begin{equation}
w_1=-\dfrac{1}{\eta_0(x)} \nabla_x\cdot(\vartheta w_0)+w_0a(R_0)\vartheta\cdot \nabla R_0
\label{w1eq_1}
\end{equation}
and
\begin{equation}
W_1=0\,.
\label{w1eq_2}
\end{equation}

\noindent
From (\ref{R2eq}) we derive the following expression for $\dfrac{\beta(S)}{\omega}R_2$:
\begin{equation}
\begin{split}
\dfrac{\beta(S)}{\omega}R_2&=\dfrac{1}{\omega}\Big (\alpha (w_0,S)+l_m(N_0)\Big )M_2+\dfrac{1}{2\omega}\Big (\partial^2_{ww} \alpha (w_0,S)w_1^2+l''_m(N_0)N_1^2\Big )M_0 \\[0.3cm]
&+\dfrac{1}{\omega}\Big (\partial_w \alpha (w_0,S)w_2+l'_m(N_0)N_2\Big )M_0+\dfrac{\mu(W_0,N_0,S)}{\omega} \dfrac{Q(x)}{Q^*}R_0-\dfrac{\gamma(S)}{\omega}R_0-\dfrac{1}{\omega}\partial_t R_0\,.
\end{split}
\label{R2eq_1}
\end{equation}
Plugging it into (\ref{m2eq}) we get:
\begin{equation}\label{eq:eqq2}
\begin{split}
\partial_tm_0+\nabla_x\cdot(v m_1)&=-\lambda_0\left(m_2-\dfrac{q}{\omega}M_2\right)+\lambda_1\left(m_2^z-\dfrac{q}{\omega}M_2^z\right)+\dfrac{q}{\omega}\Big (\alpha (w_0,S)+l_m(N_0)\Big )M_2\\[0.3cm]
&+\dfrac{q}{2\omega}\Big (\partial^2_{ww} \alpha (w_0,S)w_1^2+l''_m(N_0)N_1^2\Big )M_0+\dfrac{q}{\omega}\Big (\partial_w \alpha (w_0,S)w_2+l'_m(N_0)N_2\Big )M_0\\[0.3cm]
&+\dfrac{q}{\omega}\mu(W_0,N_0,S) \dfrac{Q(x)}{Q^*}R_0-\dfrac{q}{\omega}\gamma(S)R_0-\dfrac{q}{\omega}\partial_t R_0 -\Big (\alpha (w_0,S)+l_m(N_0)\Big )m_2\\[0.3cm]
&-\Big (\partial_w\alpha (w_0,S)w_1+l'_m(N_0)N_1\Big )m_1-\Big (\partial_w\alpha (w_0,S)w_2+l'_m(N_0)N_2\Big )m_0\\[0.3cm]
&-\dfrac{1}{2}\Big (\partial^2_{ww}\alpha (w_0,S)w_1^2+l''_m(N_0)N_1^2\Big )m_0.
\end{split}
\end{equation}
Integrating with respect to $v$ we get
\begin{equation*}
\begin{split}
\partial_tM_0+\int_V\nabla_x\cdot(v m_1)dv&=\dfrac{1}{2}\Big (\partial^2_{ww} \alpha (w_0,S)w_1^2+l''_m(N_0)N_1^2\Big )M_0+\Big (\partial_w \alpha (w_0,S)w_2+l'_m(N_0)N_2\Big )M_0\\[0.3cm]
&+\mu(W_0,N_0,S) \dfrac{Q(x)}{Q^*}R_0-\gamma(S)R_0-\partial_t R_0-\Big (\partial_w\alpha (w_0,S)w_2+l'_m(N_0)N_2\Big )M_0\\[0.3cm]
&-\dfrac{1}{2}\Big (\partial^2_{ww}\alpha (w_0,S)w_1^2+l''_m(N_0)N_1^2\Big )M_0
\end{split}
\end{equation*}
\begin{equation}
\Rightarrow\quad \partial_tM_0+\int_V\nabla_x\cdot(v m_1)dv=\mu(W_0,N_0,S) \dfrac{Q(x)}{Q^*}R_0-\gamma(S)R_0-\partial_t R_0,
\label{m2eq_1}
\end{equation}
where
\begin{equation*}
\begin{split}
\int_V\nabla_x\cdot(v m_1)dv&=\int_V\nabla_x\cdot\left[\,v\left(-\dfrac{1}{\lambda_0+\alpha (w_0,S)+l_m(N_0)}\left(\nabla_x\cdot(v m_0)-\lambda_1F(Q,S)\left(\dfrac{k_1^+}{Q^*}v\cdot \nabla Q- \dfrac{k_2^+}{S_{c,0}}v\cdot \nabla S\right)m_0\right)\right)\right]dv \\[0.3cm]
&=\nabla_x\cdot\left[\int_V -\dfrac{1}{\lambda_0+\alpha (w_0,S)+l_m(N_0)}v\otimes v\,\nabla_x\left(\dfrac{q}{\omega}M_0\right)\right]dv\\[0.3cm]
&+\nabla_x\cdot\Bigg[\dfrac{\lambda_1F(Q,S)}{\omega(\lambda_0+\alpha (w_0,S)+l_m(N_0))} \int_V v\otimes v\,q(x,\hat v)dv\, \left(\dfrac{k_1^+}{Q^*}\nabla Q- \dfrac{k_2^+}{S_{c,0}}\nabla S\right)\,M_0\Bigg]\,.
\end{split}
\end{equation*}
With the notation
\begin{equation}
\mathbb D_T(x):=\dfrac{1}{\omega}\int_Vv\otimes v\,q(x,\hat v)dv=s^2\int_{\mathbb{S}^{n-1}}\theta\otimes \theta\,q(\theta,x)d\theta=s^2\mathbb V_q(x)
\label{DT}
\end{equation} and recalling that $N_0(t,x)=M_0(t,x)+R_0(t,x)$, i.e.,
\begin{equation}
N_0=\left(1+\dfrac{\alpha (w_0,S)+l_m(N_0)}{\beta(S)}\right)M_0\,,
\end{equation}
we obtain from \eqref{m2eq_1} the following macroscopic equation for $N_0(t,x)$:
\begin{equation}
\begin{split}
\partial_tN_0&-\nabla_x\cdot\left[\dfrac{1}{\lambda_0+\alpha (w_0,S)+l_m(N_0)}\nabla_x\cdot\Big(\dfrac{\beta(S)}{\beta(S)+\alpha (w_0,S)+l_m(N_0)}\mathbb D_T(x) N_0\Big)\right]\\[0.3cm]
&+\nabla_x\cdot \left[\dfrac{\lambda_1F(Q,S)\beta(S)}{\lambda_0+\alpha (w_0,S)+l_m(N_0)}\mathbb D_T(x)\dfrac{\frac{k_1^+}{Q^*}\nabla Q\,- \frac{k_2^+}{S_{c,0}}\nabla S}{\beta(S)+\alpha (w_0,S)+l_m(N_0)}N_0\right]\\[0.3cm]
&=\dfrac{\alpha (w_0,S)+l_m(N_0)}{\alpha (w_0,S)+l_m(N_0)+\beta(S)}N_0\,\Big(\mu(W_0,N_0,S) \dfrac{Q(x)}{Q^*}-\gamma(S)\Big)\,.
\end{split}
\label{N0_eq}
\end{equation}
We denote
\begin{align}
 &\varphi(w_0,N,S):=\dfrac{\beta(S)}{\beta(S)+\alpha(w_0,S)+l_m(N)}\label{coeff-phi}\\ 
 &\varrho(w_0,N,S):=(\lambda_0+\alpha(w_0,S)+l_m(N)).\label{coeff-rho}
\end{align}
Observe that, due to \eqref{w0eq}, these are, in fact, purely macroscopic quantities.\\[-2ex]

\noindent
Integrating (\ref{w2eq}) with respect to $\vartheta \in \Theta$ gives
\begin{equation*}
\partial_t W_0 +\nabla_x\cdot\int_{\Theta } \vartheta w_1d\vartheta
=\mu_W(W_0,Q)W_0,
\end{equation*}
again due to \eqref{w0eq}. Recalling \eqref{w1eq_1}, this leads to the following macroscopic equation for the density $W_0$ of endothelial cells:
\begin{equation}
\partial_t W_0-\,\nabla\cdot \Big(\mathbb{D_{EC}}\nabla W_0\Big )+ \nabla \cdot\Big(\bbchi_a(R_0)W_0\nabla R_0\Big )=\mu_W(W_0,Q)W_0,
\label{W0eq}
\end{equation}
where $\mathbb D_{EC}(x):=\dfrac{\sigma^2}{n\eta_0(x)}\mathbb I_n$ and 
$\bbchi_a(R_0):=\dfrac{\sigma^2 a(R_0)}{n}\mathbb I_n=\eta_0(x)a(R_0)\mathbb D_{EC}(x)$.\\
The two macroscopic equations obtained in \eqref{N0_eq} and \eqref{W0eq} for the evolution of the tumor and ECs, respectively, are coupled with a  PDE for proton concentration dynamics:
\begin{equation}
\partial_t S=D_S \Delta S +g(S,N_0,W_0,Q)\,,
\label{eq:Seq}
\end{equation}
with the concrete form of $g(S,N_0,W_0,Q)$ given in \eqref{eq:concrete-g}.\\[-2ex] 

\noindent
In view of \eqref{R0eq_1}, \eqref{R1eq_1}, \eqref{m1eq_1}, and \eqref{w1eq_2}, the $\eps$-correction terms for $N$ and $W$ can be left out and ignoring the higher order terms we get the following closed PDE system characterizing the macroscopic evolution of the tumor under the influence of tissue, vasculature, and acidity:
\begin{equation}
\hspace{-0.5cm}\begin{sistem}
\partial_tN-\nabla \cdot \left[\dfrac{1}{\varrho(W,N,S)}\nabla \cdot\Big(\varphi(W,N,S)\,\mathbb D_T(x) N\Big)\right]+\nabla\cdot \left[\lambda_1\dfrac{F(Q,S)\,\varphi(W,N,S)}{\varrho(W,N,S)}\,\mathbb D_T(x)\left(\dfrac{k_1^+}{Q^*} \nabla Q- \dfrac{k_2^+}{S_{c,0}} \nabla S\right)N\right]\vspace{0.3cm}\\
\hspace{1cm}= \varphi(W,N,S)\dfrac{\alpha (W,S)+l_m(N)}{\beta(S)}N\,\Big(\mu(W,N,S) \dfrac{Q}{Q^*}-\gamma(S)\Big),\vspace{0.7cm}\\
\partial_t W=\nabla\cdot \Big(\mathbb D_{EC}\nabla W\Big )- \nabla \cdot\Big( \eta_0a(R)\mathbb D_{EC}W\nabla R\Big )+\mu_W(W,Q)W
,\vspace{0.7cm}\\
\partial_t S=D_S \Delta S +g(S,N,W,Q),
\end{sistem}
\label{mac_set_neu}
\end{equation}
with $F(Q,S)$ given in \eqref{grosses-F}, the tumor diffusion tensor $\mathbb D_T$ from \eqref{DT}, and with the coefficients $\varphi $ and $\rho $ from \eqref{coeff-phi} and \eqref{coeff-rho}, respectively, whereby we used $w=\frac{W}{|\Theta|}$, in virtue of \eqref{w0eq}. This system has to be supplemented with adequate initial conditions; its deduction has been carried out for $x\in \R^n$. For the numerical simulations to be performed in Section \ref{simulation}, we will consider it to be set in a bounded, sufficiently regular domain $\Omega \subset \R^n$ and endow it with no-flux boundary conditions.\\[-2ex]

\noindent
The next section is dedicated to specifying the remaining coefficients of the above macroscopic system.

\section{Assessment of coefficients}
\label{CoefFun}

To determine the tumor diffusion tensor $\mathbb D_T(x)$ in (\ref{DT}) we need to provide a concrete form for the (mesoscopic) orientational distribution of tissue fibers $q(x,\theta)$. Several different choices are available in the literature, see e.g. \cite{Conte2020,PH13}. As in \cite{Hunt2016} we use the orientation distribution function (ODF):
\begin{equation*}
q(x,\theta)=\dfrac{1}{4\pi|\mathbb D_W(x)|^{\frac{1}{2}}(\theta^T(\mathbb D_W(x))^{-1}\theta)^{\frac{3}{2}}}\,,
\end{equation*} 
where $\mathbb D_W(x)$ is the water diffusion tensor obtained from processing the (patient-specific) DTI data. \\[-2ex]

\noindent
We choose for the macroscopic tissue density $Q(x)$ the expression proposed in \cite{EHS}: 
\begin{equation}\label{Qx}
Q(x)=1-\dfrac{l_c^3(x)}{h^3}\,,
\end{equation} 
where $h$ is the side length of one voxel of the DTI dataset and $l_c$ a characteristic length, estimated as 
\begin{equation*}
l_c(x)=\sqrt{\frac{tr(D_W(x))h^2}{4l_1}}
\end{equation*} 
with $l_1$ being the leading eigenvalue of the diffusion tensor $D_W$.\\[-2ex]

\noindent
For the rate $\alpha (w,S)$ describing the cell phenotypic switch $p\to r$ from migration to proliferation, we choose a combination of an increasing function of $w$ and a decreasing function of $S$. As explained in Subsection \ref{subsec:meso_level}, this rate is influenced by the availability of nutrient (provided by vasculature) for sustaining the processes involved in the cell cycle and by the pH of the environment; the dynamical balance of these factors decides the migratory/mitotic fate.
Recalling \eqref{w0eq_1}, we set
\begin{equation}\label{choice-alpha}
\alpha (W,S)=\alpha_0\,\dfrac{S_n^{\sigma}W/W_{c,0}}{S_n^{\sigma}+S_n^{\sigma}W/W_{c,0}}\,\dfrac{1}{1+S/S_{c,0}}\,,
\end{equation}
where $W_{c,0}$ and $S_{c,0}$ are reference values for EC density and proton concentration, respectively.\\[-2ex]

\noindent
The second rate of phenotypic switch $p\to r$ $l_m(N)$, describing the influence of a crowed environment on cell phenotype changes, is chosen as
\begin{equation}\label{choice-lm}
 l_m(N)=l_{m,0}\left(1+\tanh(N/N_{c,0}-N^*/N_{c,0})\right),
\end{equation}
where $N^*$ represents a threshold value for glioma density: when it is exceeded, the cells are not able to move anymore. The constant $N_{c,0}$ denotes a reference value for the density of (moving and proliferating) glioma cells. \\[-2ex]

\noindent
The switching rate $r \to p$ given by the function $\beta(S)$ controls the acidity-triggered motility enhancement of formerly proliferating cells. We set
\begin{equation}\label{choice-beta}
 \beta(S)=\beta_0\left[\varepsilon+(S/S_{c,0}-S_{T,1}/S_{c,0})_+\right]
\end{equation}
with $\varepsilon\ll 1$ and $( \cdot )_{+}$ the positive part.
$S_{T,1}$ is the pH threshold which, when underrun, induces the cells to switch from a proliferative to a migrative  phenotype.\\[-2ex]

\noindent
Although tumor cells can live in an environment with substantially lower pH than that for normal tissue \cite{Gatenby,Webb2011}, when it drops below a certain threshold (which in terms of proton concentration we denote by $S_{T,2}$), the cancer cells become necrotic \cite{Gatenby2006,Webb2011}. This suggests the following cell death coefficient $\gamma(S)$:
\begin{equation}\label{choice-gamma}
\gamma (S)=\gamma_0(S/S_{c,0}-S_{T,2}/S_{c,0})_+\,.
\end{equation}

\noindent
The growth rate $\mu(W,N,S)$ can be also defined in different ways and it  should be motivated by biological evidence. As, for instance, in  \cite{EHS}, we choose a logistic-like function to describe the growth self-limitation and a growth enhancement factor depending on the vascularization $W$, along with a growth-limiting one due to acidification, like that employed in \eqref{choice-alpha}:
\begin{equation}\label{choice-mu}
\mu(W,N,S)=\mu_{N,0}\,\left(1-\dfrac{N}{K_N}-c_e\dfrac{N_e}{K_{N_e}}\right)\, \frac{W}{W_{c,0}}\, \dfrac{1}{1+S/S_{c,0}}\,.
\end{equation}

\noindent
Here, $K_N$ is the tumor carrying capacity and $\mu_{N,0}$ is a constant. 
The term $-c_e\frac{N_e}{K_{N_e}}$ is related to the extension of the model described in Section \ref{sec:extension}. In particular, $c_e=0$ when we consider the evolution of system (\ref{mac_set_neu}), while $c_e=1$ when the dynamics of healthy tissue and necrotic tissues ($N_e$) are included. ${K_{N_e}}$ represents the carrying capacity for the necrotic component.\\[-2ex] 

\noindent
Similarly, for the term $\mu_W(W,Q)$ describing proliferation of ECs we take 
\begin{equation*}
\mu_W(W,Q)=\mu_{W,0}\,\left(1-\dfrac{W}{K_W}\right) \frac{Q}{Q^*}\,,
\end{equation*}
with $K_W$ and $Q^*$ the carrying capacities for ECs and healthy tissue, respectively. \\[-2ex]

\noindent
For the tactic sensitivity $\bbchi_a(R)=\eta_0a(R)\mathbb D_{EC}$ of ECs towards (gradients of) proliferating glioma cells, we need to specify the function $a(R)$ involved in the definition (\ref{W_turning_rate}) of the EC turning rate. We choose 
\begin{equation*}
a(R)=\chi_{a_0}\,\dfrac{K_N}{(K_N+R)^2}\,,
\end{equation*}
which corresponds to the rate of change of the expression $\zeta (R)=\frac{R}{R+K_N}$ representing the equilibrium of the interactions between ECs and proliferating glioma cells $R$, scaled by a constant $\chi_{a_0}$ that is used to account for changes in the turning rate per unit of change in $d\zeta /dt$. Thereby, we assume that attachment and detachment of ECs and $R$-cells happen with the same rates.\\[-2ex]

\noindent
The reaction term in the PDE for acidity dynamics is chosen as 
\begin{equation}\label{eq:concrete-g}
g(S,N,W,Q)=g_s\frac{N}{N_{c,0}}-g_d\left(\frac{W}{W_{c,0}}+\frac{Q}{Q^*}\right)S,
\end{equation}
hence it has a source term for the production of protons by the tumor, and a decay term, as the protons are buffered by healthy tissue and also uptaken by the capillary network. \\[-2ex]

\noindent
In Table \ref{parameter} we report the range of the values for the constant parameters involved in system (\ref{mac_set_neu}), as well as the references from which they were drawn. In order to simplify, we assume a constant coefficient $\eta _0$ in the turning rate (\ref{W_turning_rate}) for ECs.
Further details on the estimation of the parameters and on the nondimensionalization of (\ref{mac_set_neu}) are given in Appendix \ref{par_est} and \ref{adim_sis}.\\
\begin{table} [!h]
\begin{center}
   \begin{tabular}{c|c|c|c} \hline
   \toprule  Parameter & Description & Value (units) & Source \\
  \midrule\vspace{0.15cm}
   $\lambda_0$ & turning frequency in $\mathcal{L}_p[\lambda(z)]$ & $0.001$\,\,(s$^{-1}$)& \cite{Sidani}\\\vspace{0.15cm}
   $\lambda_1$ & turning frequency in $\mathcal{L}_p[\lambda(z)]$ & $0.001$\,\,(s$^{-1}$) &\cite{Sidani,Engwer2}\\\vspace{0.15cm}
$\alpha_0$   & phenotype switch rate $p\to r$  & $0.0001$ $(s^{-1})$  &  \cite{pennarun2005,ko1980} \\\vspace{0.15cm}
$\beta_0$  & phenotype switch rate $r\to p$  & $0.0002$\, (s$^{-1}$)& \cite{pennarun2005,ko1980}     \\\vspace{0.15cm}
$l_{m,0}$  & overcrowding switch rate $p\to r$ & $0.0005$ (s$^{-1}$) & \cite{pennarun2005,ko1980}  \\\vspace{0.15cm} 
$N^*$  & optimal tumor density value for cell movement & $0.75\cdot K_N $  & \cite{Billy,Pham2012} \\\vspace{0.15cm}
$s$ & speed of tumor cells &  $0.0084\cdot10^{-3}$ (mm$\cdot$ s$^{-1}$) & \cite{diao2019}     \\\vspace{0.15cm}
$k^+_1$ &  attachment rate of tumor cells to tissue fibers&  $0.034\,(\text{s}^{-1})$  &\cite{Lauffenburger}\\\vspace{0.15cm}
 $k_2^+$ & interaction rate between tumor cells and protons &  $0.01\,(\text{s}^{-1})$   & \cite{Lauffenburger} \\\vspace{0.15cm}
$k^-$ & detachment rate & $0.01\,\,(\text{s}^{-1})$ & \cite{Lauffenburger}\\\vspace{0.15cm}
$\mu_{N,0}$ & tumor proliferation rate &  $9.26\cdot10^{-6}$ (s$^{-1}$)    & \cite{ke2000}\\\vspace{0.15cm}
$K_N$ & tumor carrying capacity &  $\sim10^6\,(\text{cells}\cdot\text{mm}^{-3})$ &\cite{TCs_dim} \\\vspace{0.15cm}
$\gamma_0$ & acid-induced death rate for tumor cells  &  $0.19\cdot10^{-6}$ (s$^{-1}$)   &  \cite{Swanson2011}  \\\vspace{0.15cm}
$S_{T,1}$  & proton concentration threshold for $r\to p$ & $1.995\cdot10^{-7}$ (M)     & \cite{Vaupel,Webb2011}\\\vspace{0.15cm}
$S_{T,2}$  & proton concentration threshold for tumor cell death &  $3.98\cdot10^{-7}$ (M)    & \cite{Vaupel,Webb2011}     \\\vspace{0.15cm}
$\eta_0$ & turning frequency of ECs in $\mathcal{L}_w[\eta]$ & $0.001$ (s$^{-1}$) & \cite{Szabo2010}\\\vspace{0.15cm}
$\chi_{a_0}$& duration between $R$-damped EC turnings &$4.5$ (d) & \cite{Stokes1990,Stokes1991}  \\\vspace{0.15cm}
$\sigma $ & speed of ECs & $0.0056\cdot 10^{-3}$ (mm$\cdot$ s$^{-1}$) & \cite{Czirok2013} \\\vspace{0.15cm}
$K_W$ & carrying capacity for ECs &  $\sim10^6\,(\text{cells}\cdot\text{mm}^{-3})$ &\cite{ECs_dim} \\\vspace{0.15cm}
$\mu_{W,0}$ & EC proliferation rate &$0.58\cdot10^{-6}$ (s$^{-1}$)& \cite{hanahan1996,alhazzani2019} \\\vspace{0.15cm}
$D_S$ & diffusion coefficient of protons &  $0.5\cdot 10^{-3}$ (mm$^2\cdot$s$^{-1}$) & \cite{Lide}  \\\vspace{0.15cm}
$g_s $  & proton production rate & $2.2 \cdot10^{-20}\,( \text{M}\cdot \text{mm}^3\cdot(\text{cells} \cdot s)^{-1})$ &  \cite{Martin2}     \\\vspace{0.15cm}
$g_d$ & proton removal rate &  $0.8 \cdot10^{-4}$ (s$^{-1}$) &   \cite{Martin} \\
    \bottomrule
    \end{tabular}
\end{center}
\caption{Model parameters}
 \label{parameter}
\end{table}

\section{Numerical simulations}
\label{simulation}

\begin{figure}[ht!]
\centering
\includegraphics[width=\textwidth]{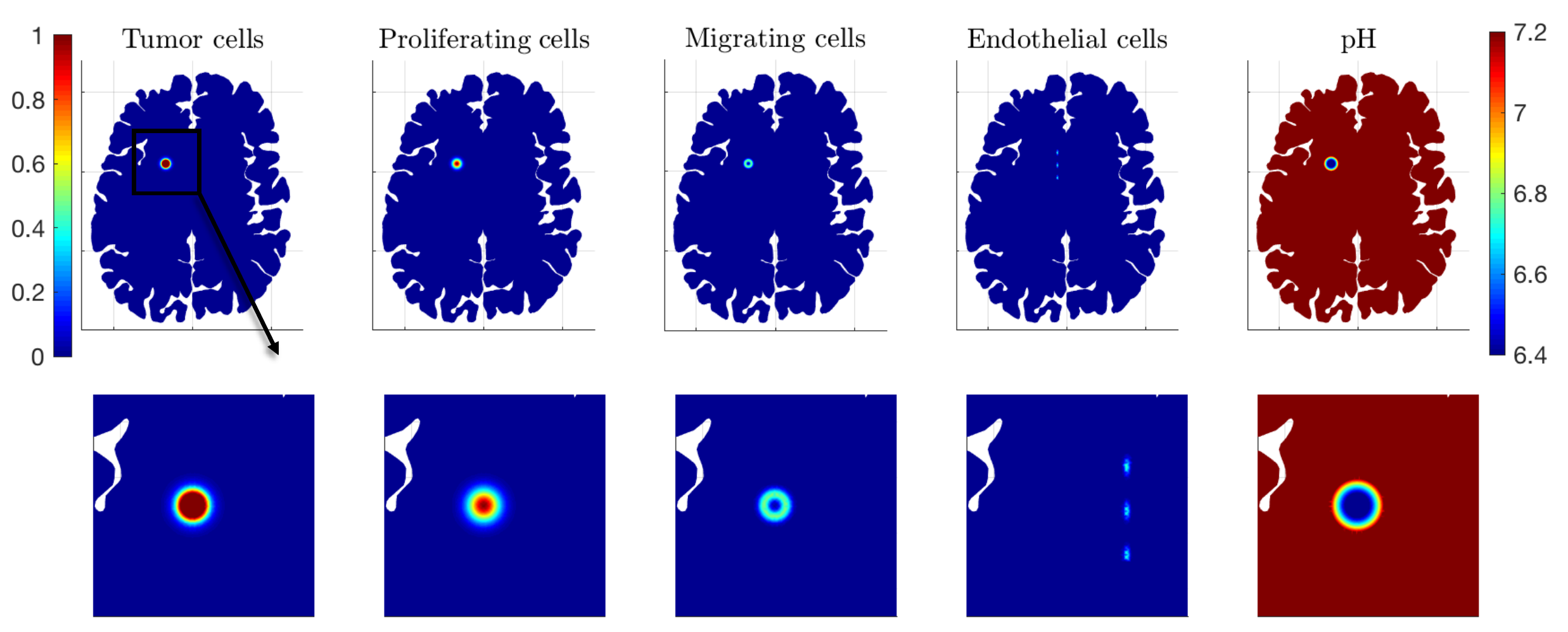}
\caption{Initial conditions for Model \eqref{mac_set_neu}.}
\label{In_Con}
\end{figure}

\begin{figure}[ht!]
\centering
\includegraphics[width=0.65\textwidth]{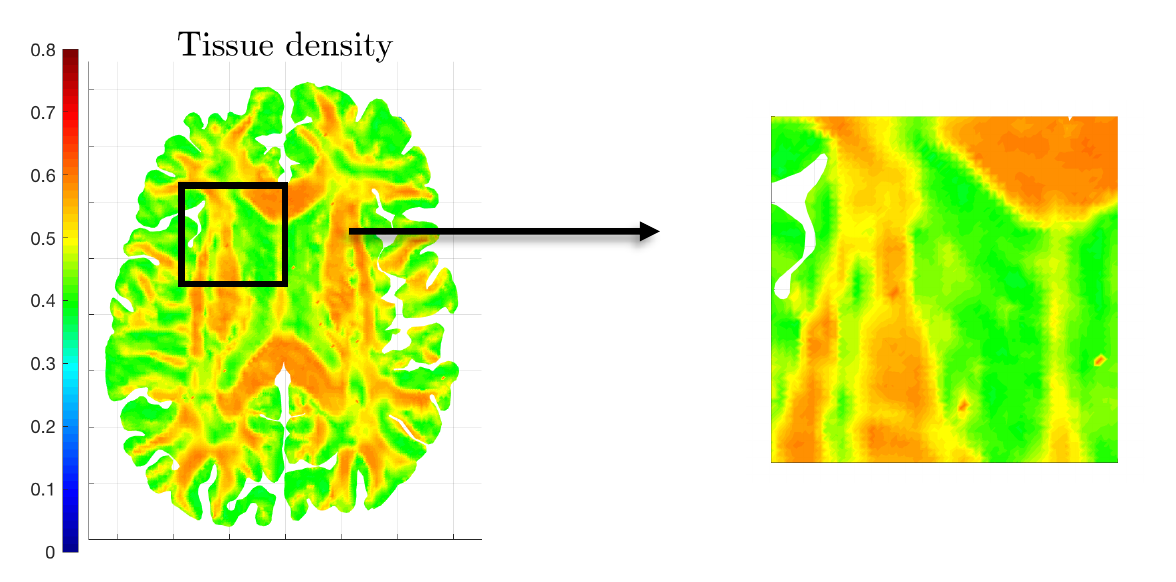}
\caption{Healthy tissue density for Model \eqref{mac_set_neu}.}
\label{Q_in}
\end{figure}

We perform 2D simulations of the resulting macroscopic system of coupled advection-diffusion-reaction equations (\ref{mac_set_neu}). For the initial conditions we take a Gaussian-like aggregate of tumor cells centered at $(x_{0,N},y_{0,N})=(-17,5)$, situated in the left-upper part of the brain slice,
\[
R_0(x,y)=e^{-\frac{\left((x-x_{0,N})^2+(y-y_{0,N})^2\right)}{8}}
\]
and a ring-like profile for the migrating tumor cells centered at the same location:
\[
M_0(x,y)=0.5\,e^{-{\left(\sqrt{(x-x_{0,N})^2+(y-y_{0,N})^2}-2\right)}^2}\,.
\]
The initial distribution of the total tumor population is given by $N_0(x,y)=R_0(x,y)+M_0(x,y)$. For the ECs we consider $x_{0,W}=-6$ and
\[
W_0(x,y)=0.5\,e^{-\frac{(x-x_{0,W})^2}{0.2}}\,\sin^6{\left(\dfrac{\pi}{8}\,y\right)} \qquad \forall\, y\in[-5, 15]
\]
reproducing the representative situation of three blood vessels close to the neoplastic region. Finally, for the acidity profile we consider a Gaussian distribution, centered at the same point as tumor cells $(x_{0,S},y_{0,S})=(-17,5)$, given by:
\[
S_0(x,y)=0.65\,e^{-\frac{\left((x-x_{0,S})^2+(y-y_{0,S})^2\right)}{10}}\,.
\]
The initial pH distribution is calculated considering that $pH_0=-\log_{10}(S_0)$. Figure \ref{In_Con} shows the plots for the initial conditions on the entire 2D brain slice, zooming then in the region $\bar{\Omega}=[-35, 5]\times[-15, 25]$. Figure \ref{Q_in} shows the initial tissue density estimated with (\ref{Qx}).\\[-2ex]

\noindent
The numerical simulations are performed with a self-developed code in Matlab (MathWorks Inc., Natick, MA). The computational domain is a horizontal brain slice reconstructed from the processing of an MRI scan. The macroscopic tensor $\mathbb D_T(x)$ is precalculated using DTI data and the ODF for the fiber distribution function (see Section \ref{CoefFun}). The dataset was acquired at the Hospital Galdakao-Usansolo (Galdakao, Spain), and approved by its Ethics Committee: all the methods employed were in accordance to approved guidelines. A Galerkin finite element scheme for the spatial discretization of the equations for tumor cells, ECs, and proton concentration is considered, together with an implicit Euler scheme for the time discretization. We present different simulations addressing several aspects.\\[-2ex]

\begin{figure}[ht!]
\centering
\includegraphics[width=\textwidth]{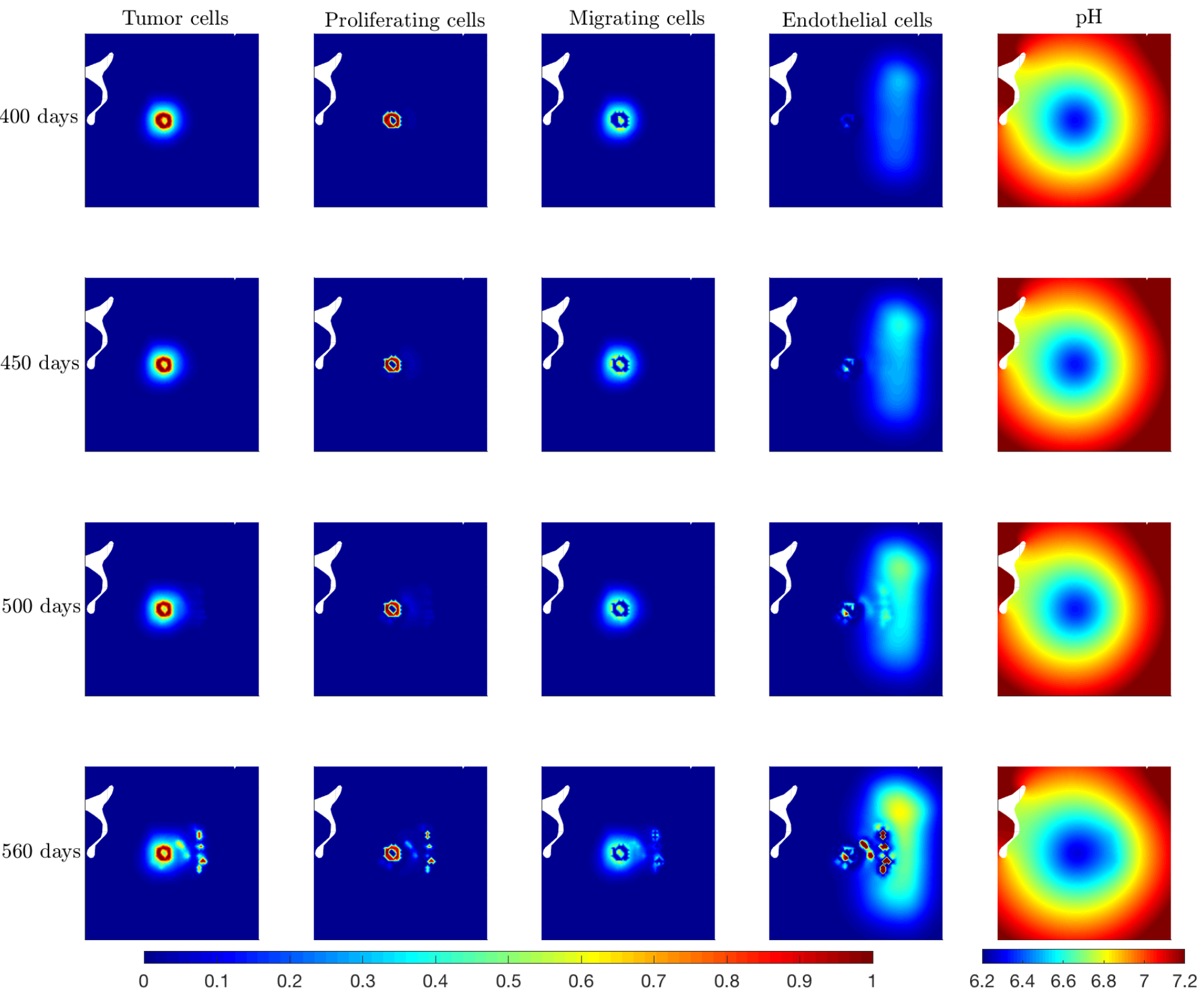}
\caption{Numerical simulation of Model (\ref{mac_set_neu}) with parameters listed in Table \ref{parameter}.}
\label{Evo_30_20_3}
\end{figure}

\noindent
Firstly, we study the behavior of species involved in \eqref{mac_set_neu} for the parameters listed in Table  \ref{parameter}. The corresponding simulation results are shown in Figure \ref{Evo_30_20_3}, where the five columns report the evolution of the  whole tumor mass ($N$), the two subpopulations of proliferating ($R$) and migrating ($M$) tumor cells, the endothelial cells ($W$), and pH (computed from $S$). The tumor spread, which seems to mainly depend on the choice of the parameters $\lambda_0$ and $s$, as well as on the EC evolution, is rather slow, with a partial exchange between the two subpopulations of tumor cells in relation to pH at the core of the tumor mass. The tumor cells increasingly adopt the proliferating phenotype when they approach ECs, as they provide the necessary nutrient and oxygen supply to sustain glioma proliferation. On the other hand, ECs diffuse and grow, with a higher accumulation around the first of the three vessels situated in the upper part of the plots and where there is more healthy tissue available to sustain their proliferation. They clearly exhibit tactic behavior toward the (pro-angiogenic growth factors released by) proliferating tumor cells mainly located around the core of the neoplasm. In particular, the subplots for the  evolution of ECs at later times (last two rows of Figure \ref{Evo_30_20_3})  show an increasing amount of high EC aggregates developing towards the tumor. This behavior can be associated with the phenomenon of microvascular hyperplasia and glomeruloid bodies. The latter are tumor-associated vascular structures that develop in the presence of high levels of VEGF and are important histopathological features of glioblastoma multiforme \cite{rojiani1996}. \\[-2ex]

\begin{figure}[ht!]
\centering
\includegraphics[width=0.8\textwidth]{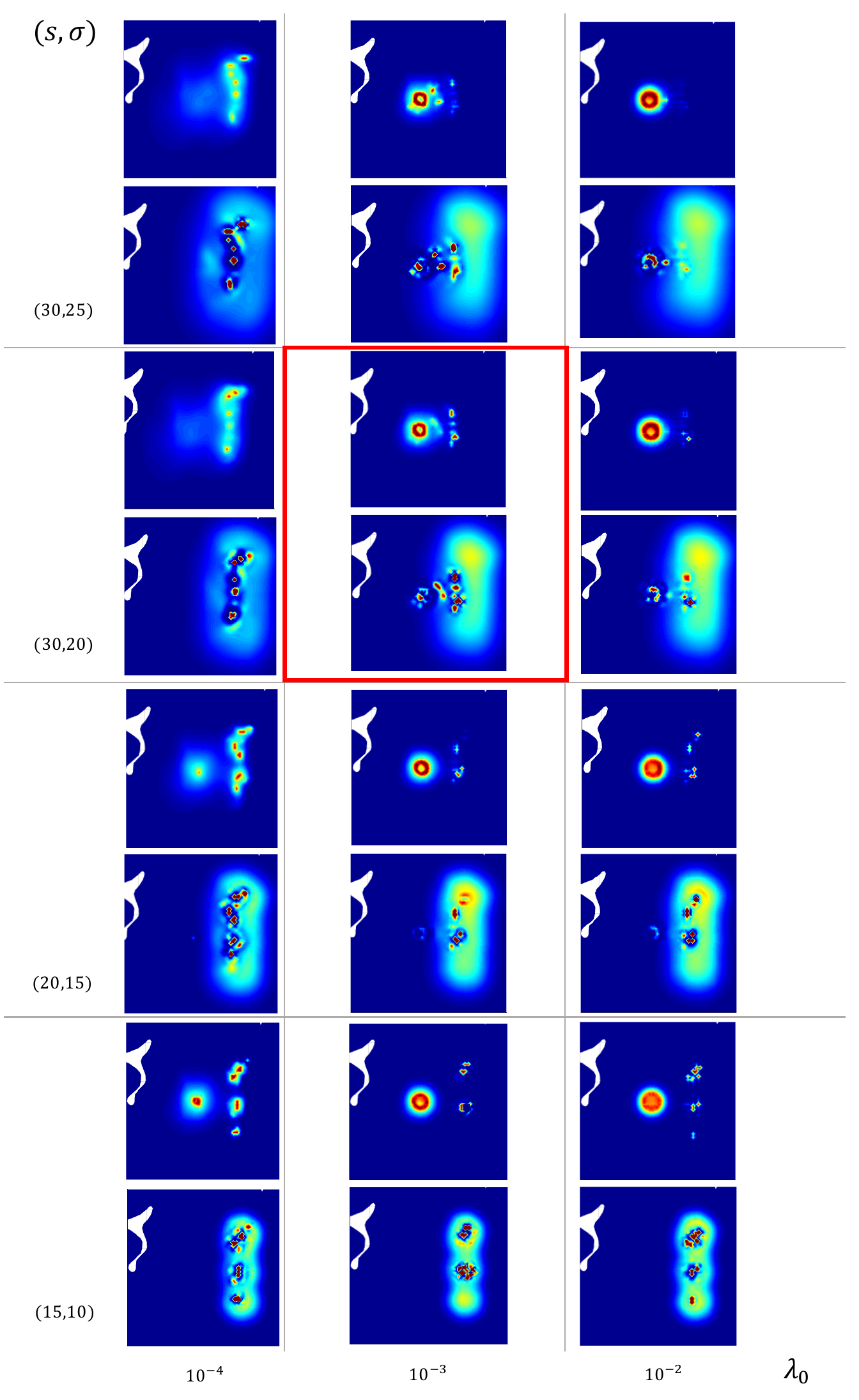}
\caption{Comparison between the evolution of tumor (first row of each box) and endothelial cells (second row of each box) for three different values of $\lambda_0$ (s$^{-1}$): $10^{-4}$, $10^{-3}$ and $10^{-2}$, and four pairs $(s,\sigma)$ of speed values ($\mu$m$\cdot$ h$^{-1}$): $(15,20)$, $(20,15)$, $(30,20)$, and $(30,25)$. All values belong to the parameter ranges reported in Appendix \ref{par_est}.}
\label{Tuning}
\end{figure}

\noindent
In Figure \ref{Tuning} we show comparisons between simulations done upon varying the parameters  $\lambda_0$, referring to the turning rate of glioma cells, and the two speeds $(s,\sigma)$ for tumor and endothelial cells, respectively, with the aim to illustrate how sensitive the model predictions are w.r.t. these parameters. The tumor and EC densities are plotted after $560$ days of evolution for three different values of $\lambda_0$ (expressed in s$^{-1}$), i.e., $10^{-4}$, $10^{-3}$ and $10^{-2}$, and for four pairs $(s,\sigma)$ of speed values (expressed in $\mu$m$\cdot$ h$^{-1}$), i.e., $(15,20)$, $(20,15)$, $(30,20)$, and $(30,25)$. The simulations suggest that vascularization at the tumor site requires a sufficiently large glioma turning rate $\lambda_0$ accompanied by relatively large EC speed $\sigma$. A too small $\lambda_0$ effects the (biologically rather unrealistic) shift of tumor cells from their original location to the site of blood vessels, where they switch to the proliferative phenotype. The faster the glioma cells are, the more pronounced is this behavior, obviously dominated by migration during its first stage and subsequent  proliferation. Increasing $\lambda_0$ by one or two orders of magnitude leads to more realistic behaviors of tumor cells and ECs, with less sensitivity towards variations in $\lambda_0$. Naturally, wide-spread hyperplasia and pronounced tumor invasion occur for higher cell speeds. \\[-2ex]

\begin{figure}[ht!]
\centering
\includegraphics[width=0.7\textwidth]{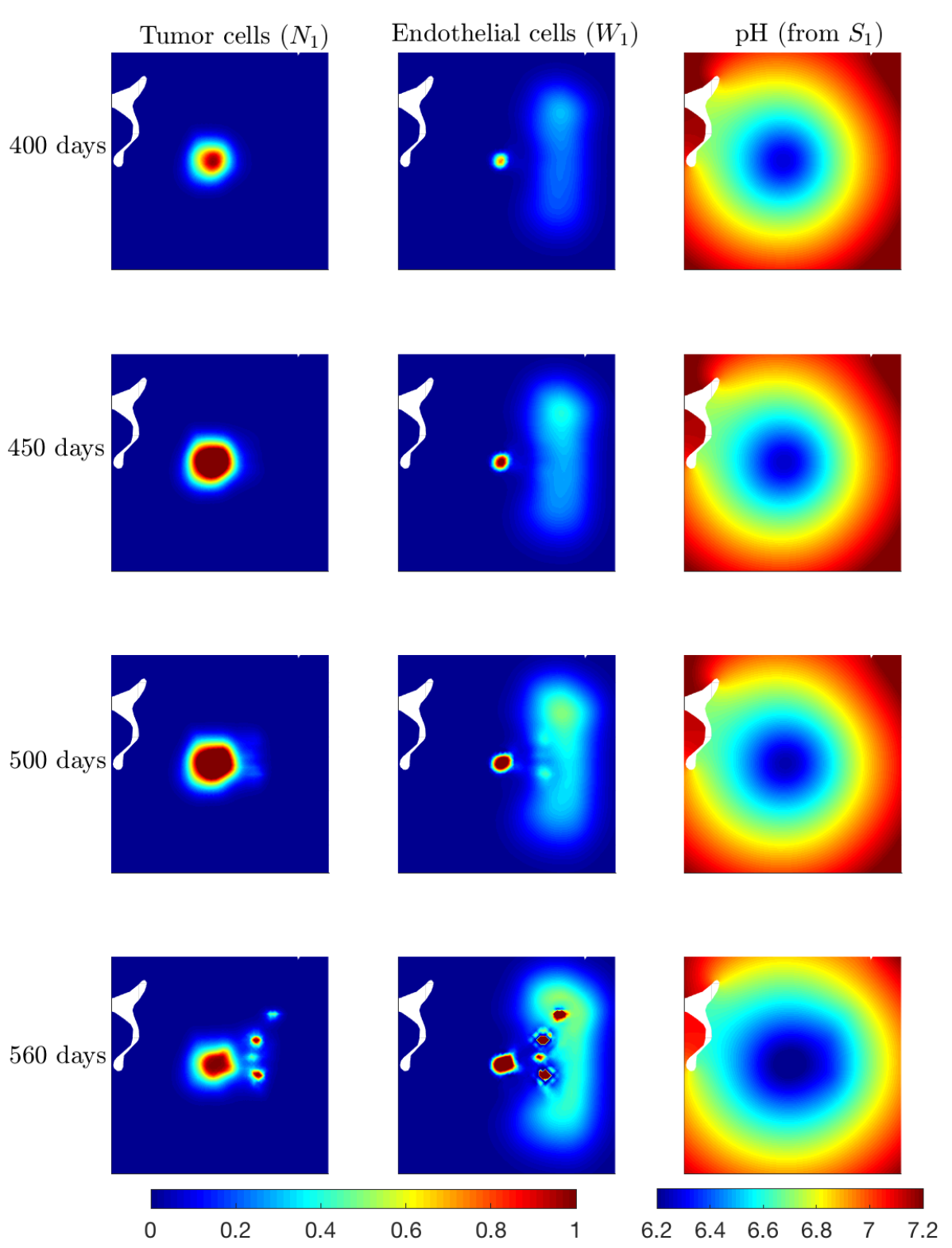}
\caption{Numerical simulation of \textit{Model NGG}, i.e., with simultaneously moving and proliferating cancer cells. The employed parameter values are listed in Table \ref{parameter}, except now $\alpha_0=0$ and $l_{m,0}=0$. }
\label{30_20_3_1pop}
\end{figure}

\noindent
In order to test the effect of the go-or-grow dichotomy on the evolution of cell populations involved in system \eqref{mac_set_neu}, we compare that model with a setting in which the tumor cells migrate and proliferate without deterring one of these phenotypes for the other. In particular, we do not differentiate between proliferating and migrating cells and accordingly let the ECs be biased by the density gradient of the whole tumor. Using a scaling argument similar to the one described in Section \ref{scaling}, we get a system of three partial differential equations for tumor cells ($N_1$), ECs ($W_1$), and protons ($S_1$), which is analogous to \eqref{mac_set_neu} with $\alpha_0=0$ and $l_{m,0}=0$. This choice of parameters reduces the former coefficient functions to $\varphi(w_0,N,S)=1$ and $\varrho(w_0,N,S)=\lambda_0$. We refer to this setting as \textit{Model NGG}. Figure \ref{30_20_3_1pop} shows the solution behavior for this new setting. The initial conditions for the three populations are the same as those shown in Figure \ref{In_Con}. Comparing Figures \ref{Evo_30_20_3} and \ref{30_20_3_1pop} we observe that the go-or-growth model predicts -as expected- a slower tumor spread, with lower cell density, that, consequently, induces lower acidity concentrations in the environment, and the differences between the two settings becoming more accentuated with increasing time. Moreover, the taxis driving ECs towards the tumor mass is stronger for the case shown in Figure \ref{30_20_3_1pop}, and accumulations of ECs indicating microvascular hyperplasia are now earlier formed and become larger.\\[-2ex]

\begin{figure}[ht!]
\centering
\includegraphics[width=\textwidth]{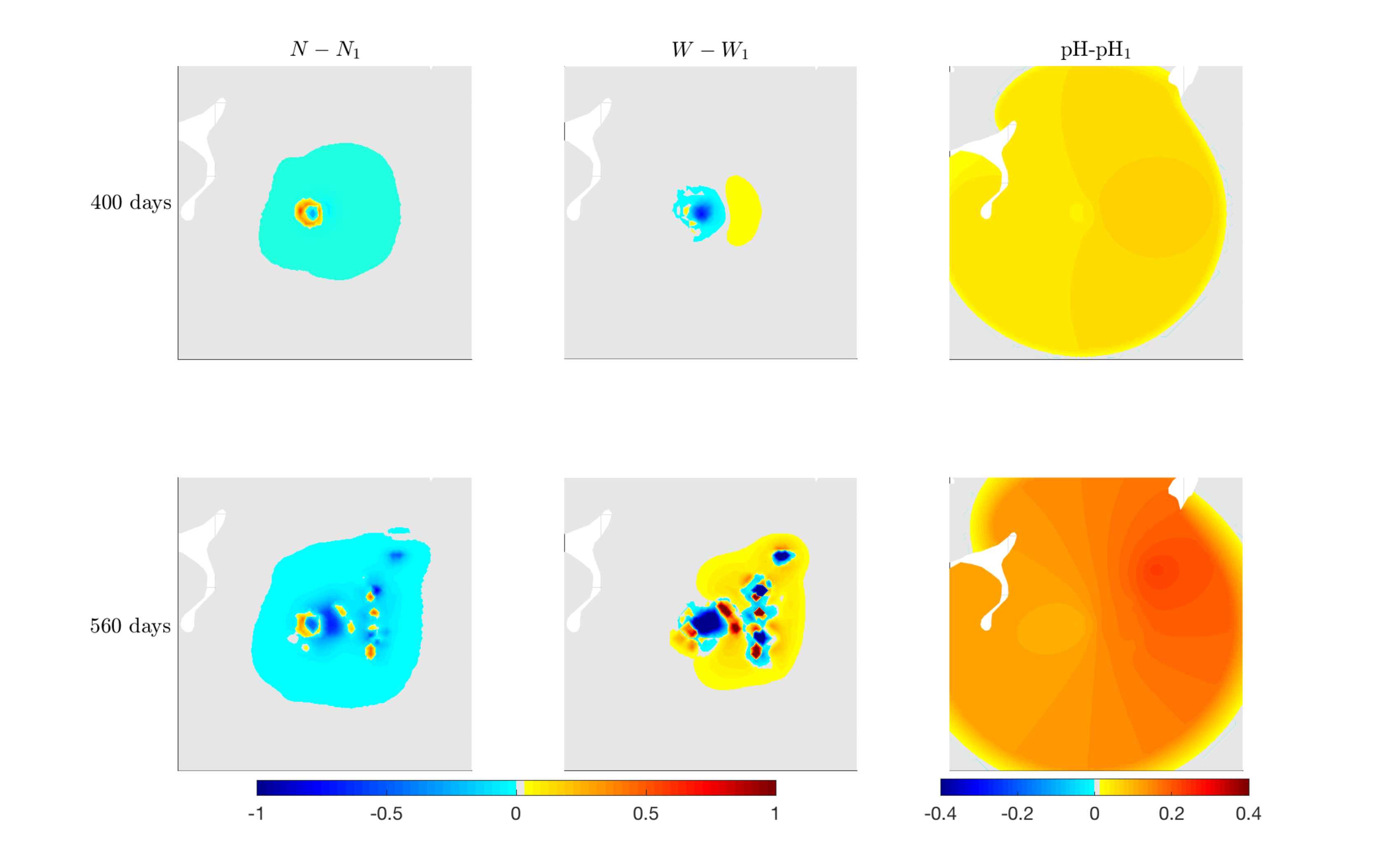}
\caption{Differences between the respective solution components of Model \eqref{mac_set_neu} and of \textit{Model NGG} at 400 (upper row) and 560 (lower row) days.}
\label{Diff_models}
\end{figure}

\noindent
To enable a direct assessment of the two settings we plot in Figure \ref{Diff_models} the differences (at 400 and 560 days) between the (overall) densities of tumor and endothelial cells for the model with go-or-grow and its {\it NGG} counterpart (the quantities for the latter are marked by the index 1), as well as between the respective pH distributions, the latter illustrated on a larger domain $\tilde{\Omega}=[-40, 10]\times[-20, 30]$. The described features concerning tumor/EC spread and aggregation along with acidity distribution can be clearly observed.

\section{Tissue degradation, necrosis, and tumor grading}
\label{sec:extension}
\noindent
We extend the above model by considering the evolution of macroscopic tissue density $Q$ and that of necrotic matter $N_e$, the latter including tissue as well as glioma cells degraded by hypoxia. The whole system consists of equations \eqref{mac_set_neu} together with the following ODEs:
\begin{equation}
\begin{sistem}
\partial_t Q=-d_Q(S)Q\vspace{0.3cm}\\
\partial_t N_e=d_Q(S)Q+\varphi(w,N,S)\dfrac{\alpha (w,S)+l_m(N)}{\beta(S)}\gamma(S)\,N,
\end{sistem}
\label{sis_ext}
\end{equation}
where the coefficient $d_Q(S)$ models the pH-triggered tissue degradation occurring when a certain acidity threshold $S_{T,Q}$ is exceeded
\begin{equation*}
 d_Q(S)=d_{0,Q}\,\left (\frac{S}{S_{c,0}}-\frac{S_{T,Q}}{S_{c,0}}\right )_{+}\,.
\end{equation*}
Thereby $d_{0,Q}>0$ denotes the tissue degradation rate and $( \cdot )_{+}$ means as usual the positive part. Details on the estimation of $d_{Q,0}$ and $S_{T,Q}$ are given in Appendix \ref{par_est}.\\[-2ex]

\begin{figure}[ht!]
\centering
\includegraphics[width=\textwidth]{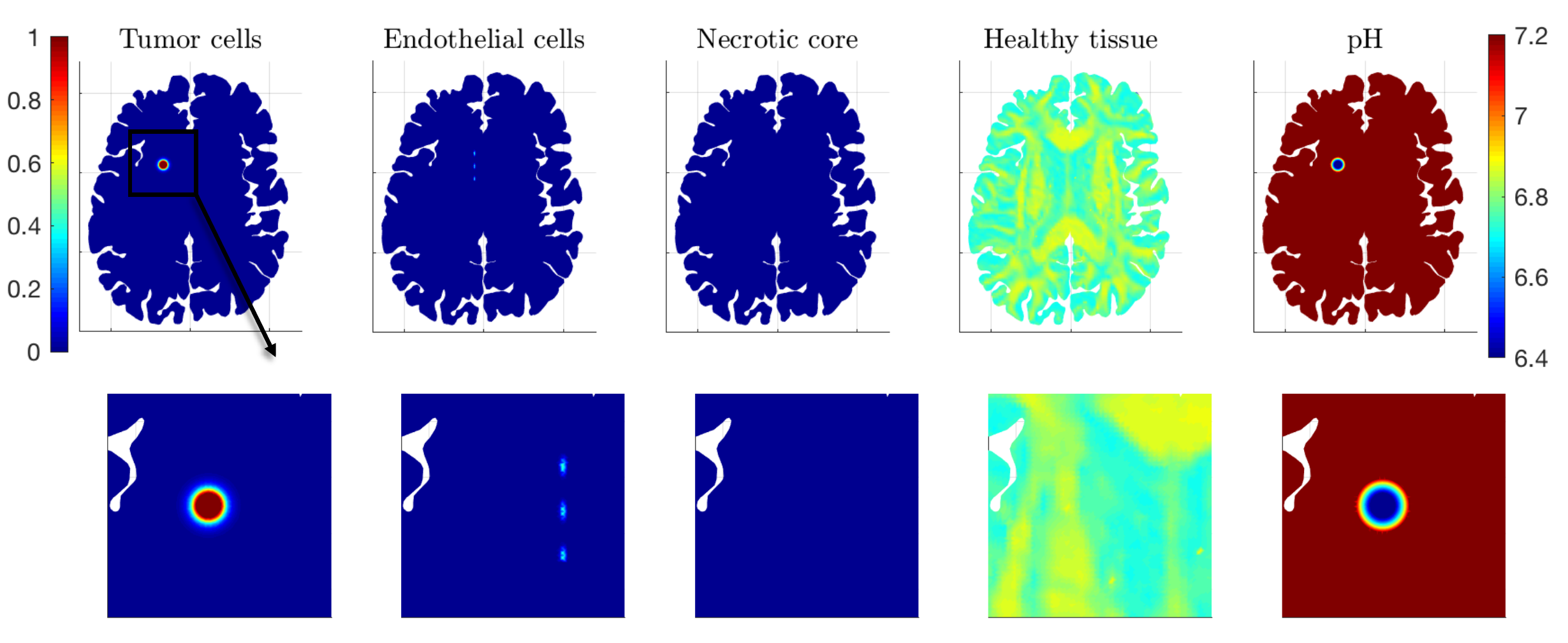}
\caption{Initial conditions for the extended Model \eqref{mac_set_neu}, \eqref{sis_ext}.}
\label{fig:In_Con_ext}
\end{figure}

\begin{figure}[ht!]
\centering
\includegraphics[width=\textwidth]{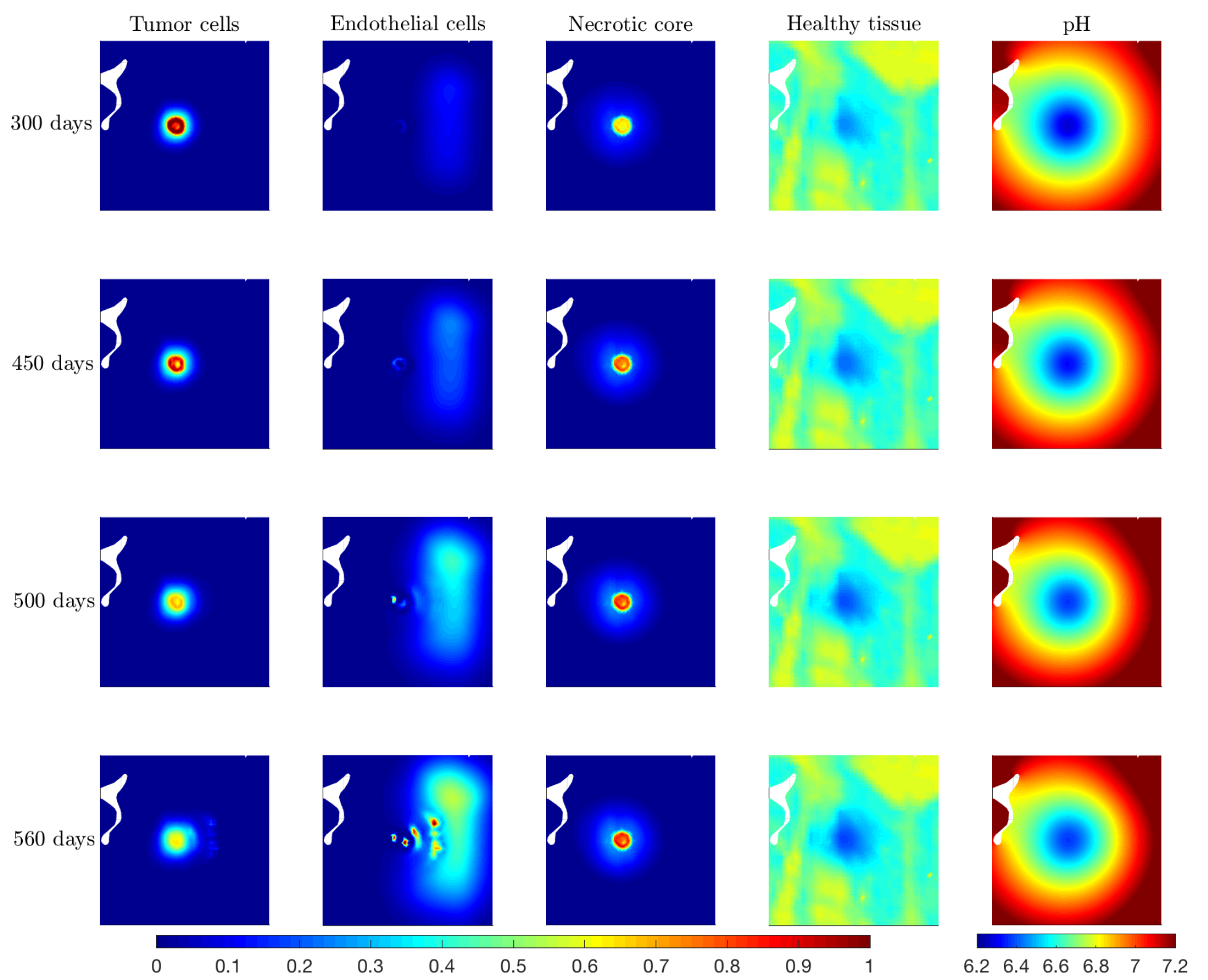}
\caption{Numerical simulation of the extended Model \eqref{mac_set_neu}, \eqref{sis_ext}. Parameters are listed in Table \ref{parameter} and the value of ECs speed is here set to $\sigma=0.0069\cdot 10^{-3}$ mm$\cdot s^{-1}$.}
\label{fig:Evo_ext_30_25_3}
\end{figure}

\begin{figure}[ht!]
\centering
\includegraphics[width=\textwidth]{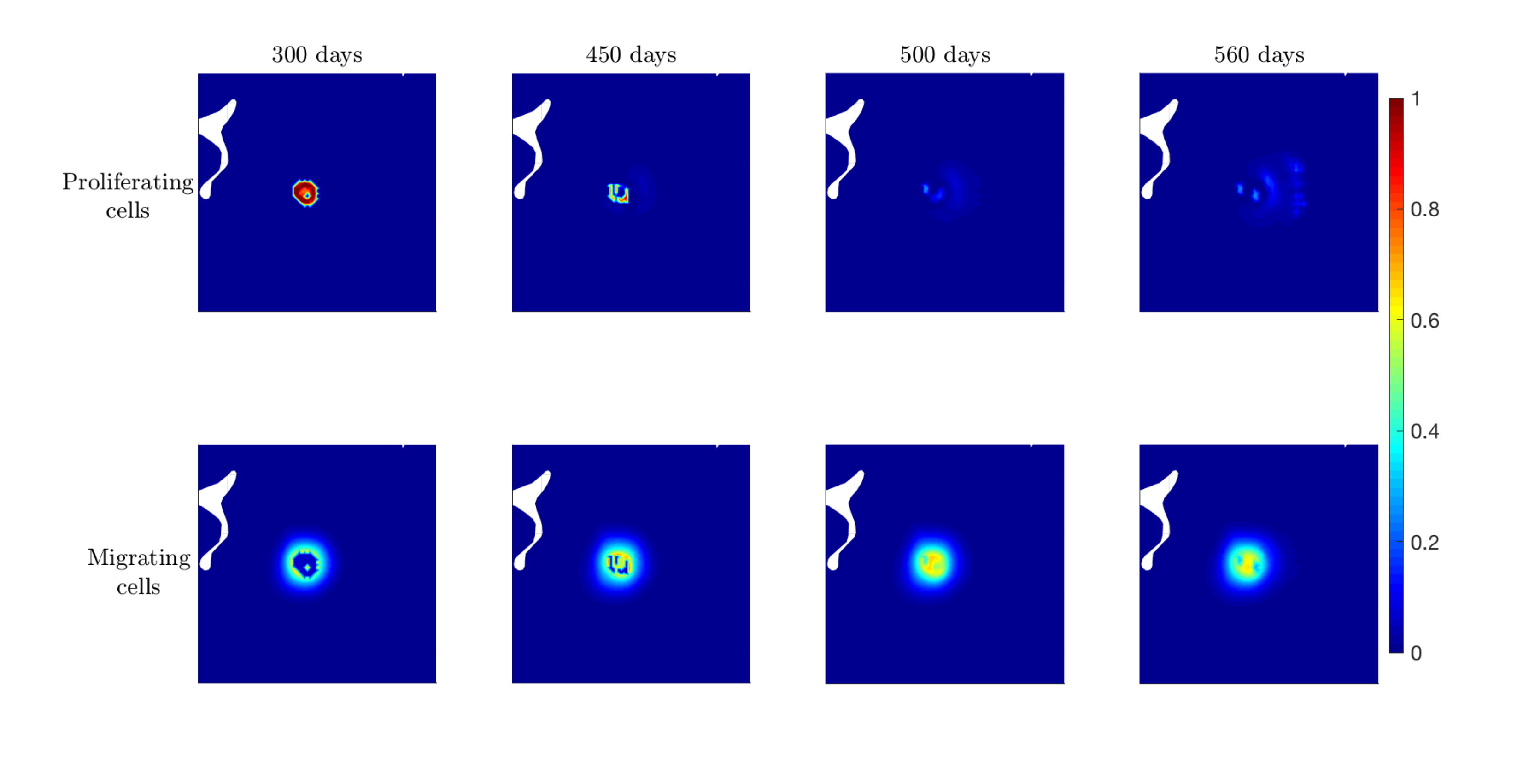}
\caption{Numerical simulation of the extended Model \eqref{mac_set_neu}, \eqref{sis_ext}. Here the evolution of the proliferating and of the migrating tumor cells is shown.}
\label{fig:RM_Evo_ext}
\end{figure}

\noindent
Numerical simulations for the extended model \eqref{mac_set_neu}, \eqref{sis_ext} are shown in Figure \ref{fig:Evo_ext_30_25_3}, corresponding to the initial conditions in Figure \ref{fig:In_Con_ext}. Figure \ref{fig:RM_Evo_ext} illustrates the evolution of proliferating and migrating glioma cells. 
Although the qualitative behavior of glioma and endothelial cells is comparable with that shown in Figure \ref{Evo_30_20_3}, the degradation of tissue by environmental acidity affects both tumor and EC proliferation, as less healthy tissue is available to sustain growth. Therefore, the simulations show lower densities for both species and, particularly, tumor cell growth is affected by the reduction of vasculature and the depletion of healthy tissue, as clearly shown by the evolution of the proliferating tumor cells in Figure \ref{fig:RM_Evo_ext} (first row). Moreover, Figure \ref{fig:Evo_ext_30_25_3} shows that lower tumor density effects less proton extrusion, i.e. higher pH.\\[-2ex]

\begin{figure}[ht!]
\centering
\includegraphics[width=0.8\textwidth]{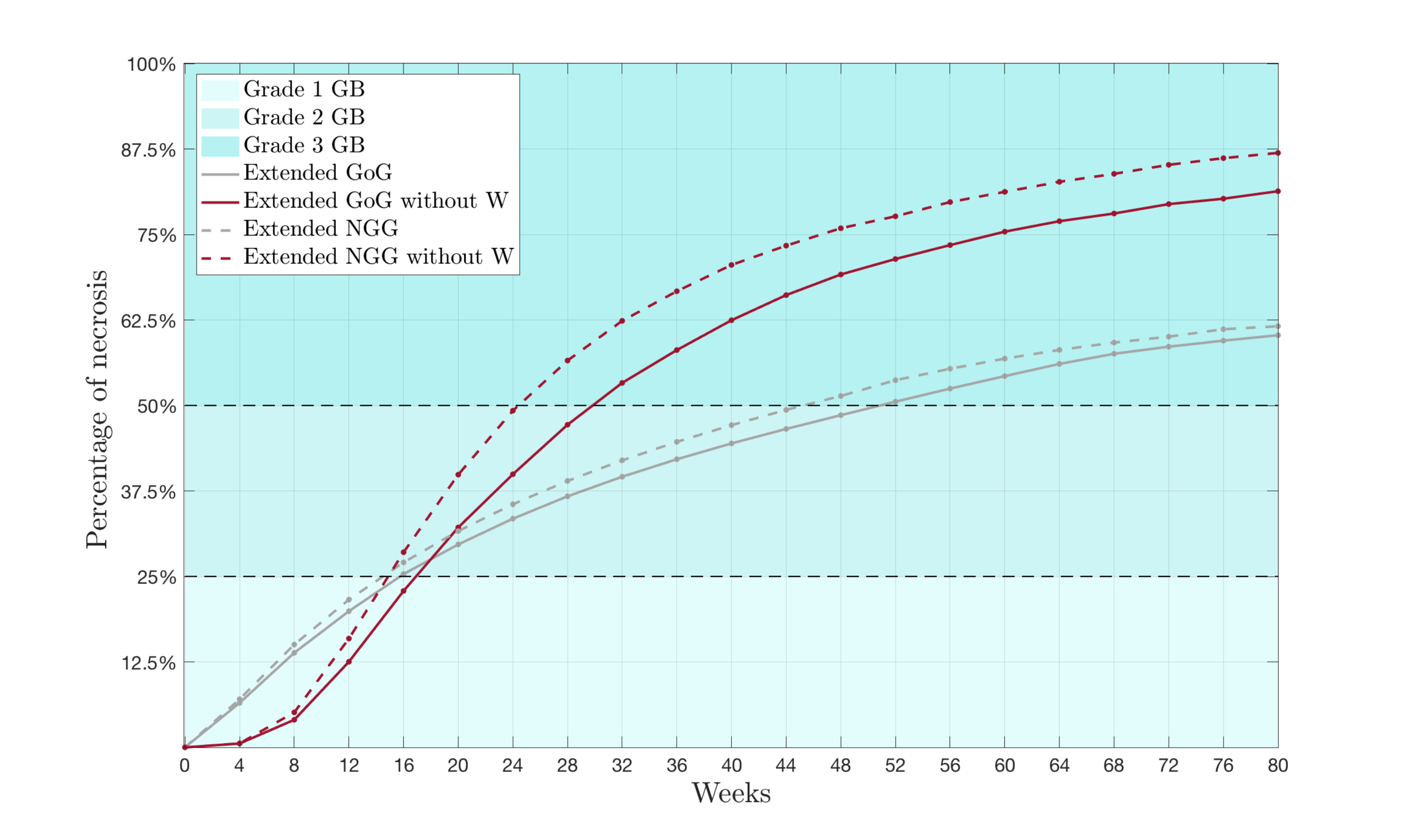}
\caption{Evolution of the grade function $G(t)$ given in \eqref{eq:tumor_grade_def}. Grey curves relate to the extended full model (i.e., including dynamics of endothelial cells) in the case with go-or-grow (solid line, model \eqref{mac_set_neu}, \eqref{sis_ext}) and without the migration-proliferation dichotomy (dotted line, \textit{Model NGG}). The red curves refer, respectively, to the same model variants, but without vascularization. 
In both cases we set $\mu_{N,0}=5.79\cdot10^{-6}$ $s^{-1}$.}
\label{fig:grading}
\end{figure}

\noindent
The above model extension enables us to perform necrosis-based tumor grading, which is essential for assessing patient survival and treatment planning. Other indicators of tumor aggressiveness are employed as well (e.g. histological patterns \cite{Wippold2037} or tumor size \cite{Pignatti2002}), however, we focus here on grading by the amount of necrosis relative to the whole tumor volume, in view of \cite{Hammoud1996,Henker2017}, where the tumor volume by itself was found to have no influence on overall survival. Following \cite{CEKNSSW}, we define the time-dependent grade $G(t)\in [0,1]$ of the simulated tumor via:
\begin{equation}
G(t):= \frac{V_{Ne}(t)}{V_{Ne}(t) +V_{N}(t)}
\label{eq:tumor_grade_def}
\end{equation}
where $V_{Ne}(t)$ and $V_{N}(t)$ denote, respectively, the fractions of necrosis and living cell densities in the visible tumor volume. They are defined as the integrals of the densities $N_e$ and $N$ over the domain defined by the level sets of the tumor population for a detection threshold of $80\%$ of the carrying capacity, which corresponds to the detection threshold for T1-Gd images \cite{Swanson2011}. We represent in Figure \ref{fig:grading} the time evolution of $G$, guided by the percentage classification in \cite{Hammoud1996}, i.e., $0<G<25\%$: grade 1, $25\%\le G< 50\%$: grade 2, and  $G\ge 50$: grade 3. The highest grade corresponds to the most aggressive tumor and the poorest survival prognosis. In particular, we compare the effect of four different scenarios on necrosis-based tumor grading: the grey curves therein refer to the extended model \eqref{mac_set_neu}, \eqref{sis_ext}) involving vascularization, i.e. EC density $W$ (solid line: go-or-grow, dotted line: {\it Model NGG}), while the red curves illustrate the evolution of $G$ for the corresponding variants of the extended model without EC dynamics (i.e., without $W$).

\noindent
Figure \ref{fig:grading} provides rich information. On the one hand, it shows that assuming the go-or-grow dichotomy leads to slower progression of neoplasia, due to the cells deterring one phenotype for the other. Indeed, in the long run, the full go-or-grow model with vascularization predicts a slower advancement towards high tumor grades. When EC dynamics are accounted for, the differences between the model with or without go-or-grow are rather small; when $W$ is, instead, not included, then such differences increase. The vascularization seems to have a significant impact: if we focus e.g., on \textit{Model NGG} (dotted curves) we see that during the first simulated 14-15 weeks the vascularization ensures a higher percentage of necrosis, after which the situation reverses, with differences becoming larger while time is advancing. The early phase (which is supposed to correspond to a lower tumor grade) might seem somehow paradoxical when thinking about blood capillaries buffering the acidity and reducing necrosis. It is, however, well-known that a tumor usually develops angiogenesis when it has reached a more advanced phase in its development and begun to get increasingly hypoxic, which leads to enhanced VEGF expression and capillary formation. The small amount of vessels prior to such stage is on the one side supporting the growth of tumor cells, while on the other side it is not able to buffer the ever increasing proton concentration triggered by the exuberant growth. Without a substantial enhancement of angiogenesis, the tumor will develop a larger necrotic component, thus receiving a higher grade. Therefore, omitting the dynamics of endothelial cells from the model might overestimate the tumor growth and spread, which for brain tumors can have a significant therapeutic impact.

\section{Discussion}
\label{conclusion}
To our knowledge this is the first continuous mathematical model with pH- and vasculature-induced phenotypic switch between moving and proliferating cells where the two cell types are seen as distinct tumor subpopulations evolving under mutual, direct and indirect interactions. As such, it can be seen as a further development of the models in \cite{Engwer,Hunt2016}. This novel model not only includes on the macroscopic scale nonlinear, myopic self-diffusion and multiple taxis (haptotaxis, repellent pH-taxis) for one of the tactic populations (glioma cells), but also features taxis of the second population (ECs) towards only one part of the former: the ECs are following, in fact, the gradient of proliferating glioma. Thereby, the chemorepellent is produced by the population (glioma cells) which tries to avoid it, but at the same time is degraded by the other tactic actor (ECs), which is, in turn, directionally biased by (a part of) the first one.
This renders the repellent acidity taxis of glioma cells both direct and indirect. The extended model involving dynamics of tissue and necrotic matter involves, too,  an indirect kind of haptotaxis, with yet more complex couplings. These features raise several highly interesting questions related to rigorous well-posedness and long term behavior of solutions. Among others, the simulations showed that for certain parameter combinations the solution can infer singularities. For a recent review of available models with multiple taxis we refer to \cite{KSSSL}. \\[-2ex] 

\noindent
We used a multiscale approach starting from single cell dynamics for glioma interacting with (macroscopically represented) extracellular acidity and tissue, we wrote the corresponding kinetic transport equations for glioma and EC density distribution functions on the mesoscopic scale, and employed a parabolic scaling to deduce the population level behavior for the model variables: glioma density, acid concentration (or pH), EC density, and -for the extended model- densities of normal tissue and necrotic matter. Thereby, tumor cell migration is driven as in previous works, e.g,  \cite{Engwer2,Engwer,PH13}, by a myopic diffusion term involving a tumor diffusion tensor which takes into account the local tissue structure. Moreover, taxis terms carrying information from the microscopic scale direct glioma migration towards increasing tissue gradients and away from highly acidic regions. We considered a heterogeneous tumor and used the go-or-grow dichotomy asserting that glioma cells can either move or proliferate, the respective behavior being transient and switching according to nutrient availability (supplied by vasculature), pH (determined by proton production and buffering), as well as crowded environments. \\[-2ex] 

\noindent
The deduced macroscopic motion of ECs is characterized, too, by a combination of diffusion and drift terms. In order not to increase too much the number of solution components, we described rather indirectly the response of ECs to pro-angiogenic growth factors (e.g. VEGFs), upon biasing their tactic motion towards proliferating tumor cells. This was realized in an alternative manner to that by which taxis terms have been obtained for glioma cells.\\[-2ex] 

\noindent
The extension of the model by taking into account the macroscopic evolution of healthy tissue and necrotic matter opened the way for necrosis-based tumor grading, which is highly relevant for diagnosis and therapy planning. A forthcoming work will be dedicated to several aspects of the latter issue.\\[-2ex] 

\noindent
The validation of model predictions through comparison with adequate patient data would be a future task to address. Although the available technology is potentially able to provide a large amount of the needed quantitative information, the standard clinical imaging and therewith associated treatment does not provide such data sets, among others due to prohibitively high costs. With the advent of technological development and the advancement of personalized medicine, however, such data may become available in the near future.

\section*{Acknowledgements}
The authors thank Prof. Juan Soler for the helpful discussions and his valuable advices on modelling issues. The authors also thank Dr. Marian Gomez Beldarrain from the Department of Neurology at the Galdakao-Usansolo Hospital (Galdakao, Spain) for the DTI data. MC acknowledges funding by the Basque Government through the BERC 2018- 2021 program and by the Spanish State Research Agency through BCAM Severo Ochoa excellence accreditation SEV-2017-0718. This project has received funding from the European Union’s Horizon 2020 research and innovation programme under the Marie Sk\l{}odowska-Curie grant agreement No. 713673. The project that gave rise to these results received the support of a fellowship from ”la Caixa” Foundation (ID 100010434). The fellowship code is LCF/BQ/IN17/11620056. CS was supported by the Federal Ministry of Education and Research BMBF, project \textit{GlioMaTh} 05M2016.

\appendix
\section{Parameter assessment}
\label{par_est}
 
\subsection{Turning rates and diffusion related parameters: $\lambda_0$, $\lambda_1$, $s$, $\sigma$, $\eta_0$, $D_S$}

$\lambda_0$, $\lambda_1$: In \cite{Sidani} the authors presented experiments on the migratory behavior and turning frequency of metastatic cancer cells from rat mammary adenocarcinoma cell line, reporting values for $\lambda_0$ in the range $[0.01 - 0.1]$ s$^{-1}$. Considering the highly aligned brain structure which influences cell migration upon enhancing cell persistence in the favorable direction of motion, we assume a reduction of the turning likelihood and take the range $[10^{-4}-10^{-2}]$ s$^{-1}$ for the parameter $\lambda_0$. The choice of $\lambda_1$ is, unfortunately, rather imprecise, since we found no data or references. In \cite{Engwer2} variations of this parameter by $\pm 50 \%$ were tested in a similar context. As proposed there, we consider the same order of magnitude for $\lambda_0$ and $\lambda_1$.\\[-2ex] 

\noindent
$s$: Different references are available for the tumor cell speed. In a recent work \cite{diao2019},  four types of typical GBM cell lines were cultured in a microfabricated 3D model to study their in vitro behavior; according to that, we consider for glioma cell speed the range $[10.2 - 30]$ $\mu$m $\cdot$ h$^{-1}$. A further upper limit for this parameter can be found in \cite{Prag}, where a maximum speed of $54-60\,\mu$m $\cdot$ h$^{-1}$ was reported for glioma cells.\\[-2ex] 

\noindent
$\sigma$: For the average speed of ECs, in \cite{Czirok2013} the range $\sigma \in [10 - 50]$ $\mu$m $\cdot$ h$^{-1}$ was reported for individual cells in several culture conditions. In vivo, the registered mean speed for motile endocardial and endothelial cells is of approximately $20$ $\mu$m $\cdot$ h$^{-1}$.\\[-2ex] 

\noindent
$\eta_0$: In \cite{Szabo2010}, the authors analyzed the statistical properties of the random streaming behavior for endothelial cell cultures. In particular, they estimated a needed time $T= 5$m to alter cell polarity and influence the cell turning. With this value, they obtained a cell speed estimation in the range of $20–40\,\mu$m $\cdot$ h$^{-1}$ within monolayers, in agreement with their empirical data about cell speed in $[10 - 30]$ $\mu$m $\cdot$ h$^{-1}$. Following their results we consider $\eta_0 \in [0.0001 - 0.003] $ s$^{-1}$.\\[-2ex] 

\noindent
$D_S$: From \cite{Lide}, we get an estimation of the diffusion coefficient for different ions; an average value between $[0.3 - 10]\cdot 10^{-3}$ mm$^2\cdot$s$^{-1}$ is reported. We test the model for different values of this parameter and we chose $D_S=0.5 \cdot 10^{-3}$ mm$^2\cdot$ s$^{-1}$. $D_S$ was rescaled in the simulations in order to account for the fast dynamics characterizing proton evolution.

\subsection{Phenotypic switch related parameters: $\alpha_0$, $\beta_0$, $l_{m,0}$, $N^*$, $S_{T,1}$, $S_{T,2}$, $\gamma_0$}

$\alpha_0$, $\beta_0$, $l_{m,0}$: The estimation of reasonable ranges for the values of the phenotypic switch parameters $\alpha_0$ and $\beta_0$ might be quite imprecise, since there are no specific data or references available. However, considering the experiment done in \cite{pennarun2005,ko1980}, it is possible to define a wide range for the duration of the glioma cell cycle that translates into $\alpha_0,\beta_0 \in [0.09 - 1] \cdot 10^{-4}$ s$^{-1}$. Analogous arguments apply for the rate $l_{m,0}$, for which there are not estimations directly derived from biological data. For these reasons we test the model for several values of $\alpha_0$, $\beta_0$, $l_{m,0}$ with order of magnitude $10^{-4}$.\\[-2ex] 

\noindent
$N^*$: Due to the lack of biological data, for the estimation of this parameter related to the total tumor density level that still allows cells movement, we choose it to be proportional to the estimated tumor carrying capacity, referring to the range $[0.6-0.9]\cdot K_N$ proposed in \cite{Pham2012,Billy}.\\[-2ex] 

\noindent
$S_{T,1}$, $S_{T,2}$: Following the pH range $[6.4-7.3]$ proposed in \cite{Vaupel, Webb2011} for the brain tumor microenvironment, we choose the thresholds values that determines the phenotypic switch from a proliferating to a migrating cells ($S_{T,1}$) and the acid-mediated death of the resting cells ($S_{T,2}$). We set $S_{T,1}=1.995\cdot10^{-7}$ M (referring to a pH$=6.7$),  and $S_{T,2}=3.98\cdot10^{-7}$ M (referring to a pH$=6.4$).\\[-2ex] 

\noindent
$\gamma_0$: In \cite{Swanson2011}, the authors assumed the cell necrosis rate to be proportional to their metabolic rate $\mu_{N,0}$. In particular, this estimation appears reasonable, considering also our assumption in the parabolic scaling procedure about having similar time scales for birth and death. Therefore, we set $\gamma_0=\mu_{N,0}/50$.

\subsection{Adhesion related parameters: $k_1^+$, $k_2^+$, $k^-$, $\chi_{a_0}$}

$k_1^+$, $k_2^+$: For the estimation of the attachment rates between tumor cell and ECM or protons, we refer to \cite{Lauffenburger}. In particular, for both the cell-ECM attachment rate and the cell-protons interaction rate, we set $k^+_1=k_2^+=10^{4}\,(\text{M\,s})^{-1}$. Then, assuming that the main ECM component is collagen, with a molecular weight of $\approx 300\text{kDa}$, and taking into account the reference value $Q^*$ (in Table \ref{ref_val}), we deduce $k_1^+=0.034\,\text{s}^{-1}$. Analogously, considering the reference value for the protons concentration $S_{c,0}$ (in Table \ref{ref_val}), we get $k_2^+=0.01\,\text{s}^{-1}$.\\[-2ex] 

\noindent
$k^-$: For the estimation of the cells-ECM and cells-protons detachment rate, referring to \cite{Lauffenburger}, we set $k^-=0.01\,\text{s}^{-1}$.\\[-2ex] 

\noindent
$\chi_{a_0}$: We estimate the parameter $\chi_{a_0}$ by considering the values reported in \cite{Stokes1990} and \cite{Stokes1991} for the chemotactic sensitivity. In particular, in \cite{Stokes1990} the authors analyzed the chemotactic coefficient of migrating endothelial cells in gradients of aFGF, measuring a maximum chemotactic response of $2600$ cm$^2\cdot$(M s)$^{-1}$ at a concentration of aFGF around $10^{-10}$ M. Instead, in the further work \cite{Stokes1991}, the authors analyzed the changes of this parameter in response to cell speed and persistence time. Taking into account the above described range for the EC speed, we get $\chi_{a_0}\in[3.09 - 4.5]$ d.

\subsection{Proliferation related parameters: $\mu_{N,0}$, $K_N$, $\mu_{W,0}$, $K_W$}     

$\mu_{N,0}$: For the estimation of glioma growth rate, we analyze the doubling times reported in \cite{ke2000} for several glioma cell lines. There, the authors reported a range of variably between $21.1$h and $46$h, which translated into  $\mu_{N,0}\in[0.42 -0.9]\cdot10^{-5}$ s$^{-1}$. \\[-2ex] 

\noindent
$K_N$, $K_W$: Considering that the mean diameter of a glioma cell is around $[12-14]\,\mu\text{m}$ \cite{TCs_dim}, we estimate a value for the tumor carrying capacity of $K_N\sim10^6\,\text{cells}\cdot\text{mm}^{-3}$. In the same way, considering a mean diameter for an endothelial cell of $[10-20]\,\mu\text{m}$ \cite{ECs_dim}, we set $K_W\sim10^6\,\text{cells}\cdot\text{mm}^3$.\\[-2ex] 

\noindent
$\mu_{W,0}$: The doubling time of EC density has been estimated in several experiments to vary between the different phases of an endothelial colony growth until the formation of a monolayer. In \cite{hanahan1996,alhazzani2019}, the authors gave a range of variability for the value of the EC doubling time of $[3 - 13]$ d. This leads to $\mu_{W,0} \in [0.62 - 2.7]\cdot10^{-6}$ s$^{-1}$.

\subsection{Production and consumption related parameters $g_{0,N}$, $g_{0,W}$, $d_{0,Q}$, $S_{T,Q}$}

$g_s$: In \cite{Gatenby}, the authors estimated the rate of proton production due to tumor cell activity by fitting their equation for proton dynamics (analogous to our PDE \eqref{eq:Seq}) to a converted form of the data in \cite{Martin2}. In particular, in \cite{Martin2} pH measurements were taken at a variety of points within both the tumor and surrounding healthy tissue for four composite cases, giving a geometric mean for the production rate of $2.2\cdot 10^{-20}$ M $\cdot$ mm$^3\cdot (\text{s}\cdot\text{cell})^{-1}$.\\[-2ex] 

\noindent 
$g_d$: Following \cite{Martin,Gatenby}, for the rate of proton uptake by vasculature we consider the range of variability given by $[0.66 - 1.1]\cdot 10^{-4}$ s$^{-1}$. As for $D_S$, proton production and consumption rates were rescaled in the simulations in order to account for the fast dynamics characterizing proton evolution.\\[-2ex] 

\noindent
$d_{0,Q}$: For the rate of tissue degradation due to the acidic environment, in \cite{Swanson2011} the authors proposed an estimation choosing the parameter such that $10\%$ necrosis gives tissue a $50$-day half-life. Starting from the value proposed in \cite{Swanson2011}, we test a wider range of possible estimations, that translates into $d_{0,Q}\in[0.005 - 0.07]$ d$^{-1}$. \\[-2ex] 

\noindent
$S_{T,Q}$: In \cite{Vaupel}, tissue pH values in normal brain and in brain tumors were reported. Specifically, considering that these values vary depending on the type of brain tissue (i.e., gray matter, white matter, cerebellum), the minimum pH requires for the normal cell activity is in the range of $[6.94 - 6.74]$. For these reasons, we set $S_{T,Q}=1.995\cdot 10^{-7}$ M (referring to a pH$=6.7$).

\section{Nondimensionalization}
\label{adim_sis}
To proceed with nondimensionalizing the system, we firstly observe that the variables $N$, $W$, and $N_e$ involved in systems (\ref{mac_set_neu}) and (\ref{sis_ext}) are expressed in cells/mm$^3$, $Q$ in g/mm$^3$, while the concentration of protons $S$ is given in mol/liter (=:M). The reference values we use for the nondimensionalization are listed in Table \ref{ref_val}. In particular, we rescale the tumor, EC, and necrotic matter (dead cells and tissue) densities with respect to their carrying capacities, i.e., $N_{c,0}=K_N$, $W_{c,0}=K_W$, and $N_{e,0}=K_N$, assuming a similar carrying capacity for tumor cells and necrosis. 

\begin{table} [!h]
\begin{center}
   \begin{tabular}{c|c|c|c} \hline
   \toprule  Parameter & Description & Value (units) & Source \\
  \midrule\vspace{0.15cm}
   T & time & $1$ (d) & \\\vspace{0.15cm}
   L & length & $0.875$ (mm)& \\\vspace{0.15cm}
$N_{c,0}$  & tumor cell density & $10^{6}$ (cell$\cdot$mm$^{-3}$)   & this work \\\vspace{0.15cm}
$W_{c,0}$ & EC density &$10^{6}$ (cell$\cdot$mm$^{-3}$)& this work \\\vspace{0.15cm}
$S_{c,0}$  & Proton concentration & $10^{-6}$ (M) & \cite{Vaupel} \\\vspace{0.15cm}
$Q^*$  & healthy tissue density & $10^{-3}$ (mg$\cdot$mm$^{-3}$) & \cite{Kaufman2005} \\
$N_{e,0}$  & density of necrotic matter & $10^{6}$ (cell$\cdot$mm$^{-3})$ & this work \\
    \bottomrule
    \end{tabular}
\end{center}
\caption{Reference variables for the nondimensionalization.}
 \label{ref_val}
\end{table}

\noindent We nondimensionalize the partial differential equations introduced above as follows:
\begin{equation*}
\tilde{t}=\dfrac{t}{T},\qquad \tilde{x}=\dfrac{x}{L},\qquad \tilde{N}=\dfrac{N}{K_N},\qquad\tilde{W}=\dfrac{W}{K_W},\qquad\tilde{S}=\dfrac{S}{S_{c,0}},\quad\quad\tilde{Q}=\dfrac{Q}{Q^*},\qquad \tilde{N_e}=\dfrac{N_e}{K_N}\,.
\end{equation*}
The proper scaling of the parameters involved in the macroscopic setting then reads 

\begin{equation*}
\begin{split}
& \tilde{\alpha_0}=\dfrac{\alpha_0}{\lambda_0},\qquad\tilde{l}_{m,0}=\dfrac{l_{m,0}}{\lambda_0},\qquad \tilde{N}^*=\dfrac{N^*}{K_N},\qquad \tilde{\beta}_0=\dfrac{\beta_0}{\lambda_0}\qquad\tilde{S}_{T,j}=\dfrac{S_{T,j}}{S_{c,0}}\quad (j=1,2),\\[0.3cm]
& \tilde{k_1^+}=\dfrac{k_1^+}{\lambda_0}, \qquad\tilde{k_2^+}=\dfrac{k_2^+}{\lambda_0}, \qquad\tilde{k^-}=\dfrac{k^-}{\lambda_0},\qquad\tilde{\mu}_{N,0}=\mu_{N,0}\,T,\qquad \tilde{\gamma}_0=\gamma_0\,T,\\[0.3cm]
&\tilde{\lambda_1}=\dfrac{\lambda_1}{\lambda_0},\qquad\tilde{D}_T=\dfrac{1}{\lambda_0}\dfrac{T}{L^2}D_T,\qquad\tilde{\mathbb{D}}_{EC}=\dfrac{T}{L^2}\,\mathbb{D}_{EC},\qquad \tilde{\mu}_{W,0}=\mu_{W,0}\,T,\qquad \tilde{\chi}_{a_0}=\dfrac{{\chi}_{a_0}}{K_N},\\[0.3cm]
&\tilde{D_S}=\dfrac{T}{L^2}D_S,\qquad \tilde{g}_{s}=g_{s}\dfrac{T}{S_{c,0}},\qquad \tilde{g}_{d}=g_{d}\,T,\qquad\tilde{d}_{0,Q}=d_{0,Q}\,T,\qquad \tilde{S}_{T,Q}=\dfrac{S_{T,Q}}{S_{c,0}}
\end{split}
\end{equation*}
Dropping the tilde (" $\tilde{ }$ ") in the new variables and parameters, the differential equations in system (\ref{mac_set_neu}) keep the same form, with the following rescaled functions:

\begin{equation*}
\begin{split}
&\tilde{\alpha}(\tilde{W},\tilde{S})=\tilde{\alpha}_{0}\,\dfrac{\tilde{W}}{1+\tilde{W}}\,\dfrac{1}{1+\tilde{S}},\qquad \tilde l_m(\tilde N)=\tilde l_{m,0}(1+\tanh (\tilde N-\tilde N^*)),\qquad \tilde \rho (\tilde W,\tilde N, \tilde S)=1+\tilde \alpha (\tilde W,\tilde S)+\tilde l_m(\tilde N),\\[0.3cm] &\tilde \beta (\tilde S)=\tilde \beta _0(\varepsilon+(\tilde S-\tilde S_{T,1})_+),\qquad \tilde \varphi (\tilde W,\tilde N,\tilde S)=\frac{\tilde \beta (\tilde S)}{\tilde \beta (\tilde S)+\tilde{\alpha}(\tilde{W},\tilde{S})+\tilde l_m(\tilde N)},\qquad \tilde a(\tilde R)=\dfrac{\tilde{\chi}_{a_0}}{(1+\tilde R)^2},\\[0.3cm] &\tilde{\mu}(\tilde{W},\tilde{N},\tilde{S}):=\tilde{\mu}_{N,0}\,\left(1-\tilde{N}-\tilde{N_e}\right)\,\tilde{W}\dfrac{1}{1+\tilde{S}}, \qquad 
\tilde{\mu}_W(\tilde{W},\tilde{Q}):=\tilde{\mu}_{W,0}\left(1-\tilde{W}\right)\tilde{Q},\qquad \tilde \gamma (\tilde S)=\tilde \gamma_0(\tilde S-\tilde S_{T,2})_+.\\[0.3cm]
&\tilde g(\tilde N,\tilde S,\tilde W,\tilde Q)=\tilde g_s\tilde N-\tilde g_g(\tilde W+\tilde Q)\tilde S, \qquad \tilde{B}(\tilde{Q},\tilde{S})=\left(\tilde{k}_1^+\tilde{Q}+\tilde{k}_2^+\tilde{S}+\tilde{k}^-\right),\\[0.3cm]
& \tilde{F}(\tilde{Q},\tilde{S})=\dfrac{\tilde{k}^-}{\tilde{B}(\tilde{S},\tilde{Q})^2\left[\tilde{B}(\tilde{Q},\tilde{S})+1+\tilde{\alpha}(\tilde{W},\tilde{S})+\tilde{l}_m(\tilde{N})\right]}.
\end{split}
\end{equation*}

\addcontentsline{toc}{section}{References}
\bibliographystyle{plain}
\bibliography{bibl_new}

\begin{thebibliography}{10}

\bibitem{TCs_dim}
Estimation taken from.
\newblock
  https://bionumbers.hms.harvard.edu/bionumber.aspx{?}s=n{\&}v=0{\&}id=108941.

\bibitem{ECs_dim}
Estimation taken from.
\newblock http://www.lab.anhb.uwa.edu.au/mb140/MoreAbout/Endothel.htm.

\bibitem{Alfonso17}
J.C.L. Alfonso, K.~Talkenberger, M.~Seifert, B.~Klink, A.~Hawkins-Daarud, K.R.
  Swanson, H.~Hatzikirou, and A.~Deutsch.
\newblock The biology and mathematical modelling of glioma invasion: a review.
\newblock {\em Journal of The Royal Society Interface}, 14(136):20170490,
  November 2017.

\bibitem{alhazzani2019}
K.~Alhazzani, A.~Alaseem, M.~Algahtani, S.~Dhandayuthapani, T.~Venkatesan, and
  A.~Rathinavelu.
\newblock Angiogenesis in cancer treatment: 60 years’ swing between promising
  trials and disappointing tribulations.
\newblock {\em Anti-Angiogenesis Drug Discovery and Development: Volume 4},
  4:34, 2019.

\bibitem{Bell-KTAP}
N.~Bellomo.
\newblock {\em Modeling Complex Living Systems}.
\newblock Birkh\"{a}user Boston, 2008.

\bibitem{bellomo2010}
N.~Bellomo, A.~Bellouquid, J.~Nieto, and J.~Soler.
\newblock Complexity and mathematical tools toward the modelling of
  multicellular growing systems.
\newblock {\em Mathematical and Computer Modelling}, 51(5-6):441--451, 2010.

\bibitem{Berens1999}
M.E. Berens and A.~Giese.
\newblock {\textquotedblleft}...those left behind.{\textquotedblright}
  {B}iology and oncology of invasive glioma cells.
\newblock {\em Neoplasia}, 1(3):208--219, 1999.

\bibitem{Billy}
F.~Billy, B.~Ribba, O.~Saut, H.~Morre-Trouilhet, T.~Colin, D.~Bresch, J.-P.
  Boissel, E.~Grenier, and J.-P. Flandrois.
\newblock A pharmacologically based multiscale mathematical model of
  angiogenesis and its use in investigating the efficacy of a new cancer
  treatment strategy.
\newblock {\em Journal of Theoretical Biology}, 260(4):545--562, 2009.

\bibitem{boettger-etal-12}
K.~B\"ottger, H.~Hatzikirou, A.~Chauviere, and A.~Deutsch.
\newblock Investigation of the migration/proliferation dichotomy and its impact
  on avascular glioma invasion.
\newblock {\em Mathematical Modelling of Natural Phenomena}, 7:105--135, 2012.

\bibitem{Brat2004b}
D.J. Brat, A.A. Castellano-Sanchez, S.B. Hunter, M.~Pecot, C.~Cohen, E.H.
  Hammond, S.N. Devi, B.~Kaur, and E.G.~Van Meir.
\newblock Pseudopalisades in glioblastoma are hypoxic, express extracellular
  matrix proteases, and are formed by an actively migrating cell population.
\newblock {\em Cancer Research}, 64(3):920--927, 2004.

\bibitem{Brat2004}
D.J. Brat and E.G.~Van Meir.
\newblock Vaso-occlusive and prothrombotic mechanisms associated with tumor
  hypoxia, necrosis, and accelerated growth in glioblastoma.
\newblock {\em Laboratory Investigation}, 84(4):397--405, 2004.

\bibitem{Chouaib2012}
S.~Chouaib, Y.~Messai, S.~Couve, B.~Escudier, M.~Hasmim, and M.Z. Noman.
\newblock Hypoxia promotes tumor growth in linking angiogenesis to immune
  escape.
\newblock {\em Frontiers in Immunology}, 3, 2012.

\bibitem{Colombo2015}
M.C. Colombo, C.~Giverso, E.~Faggiano, C.~Boffano, F.~Acerbi, and P.~Ciarletta.
\newblock Towards the personalized treatment of glioblastoma: Integrating
  patient-specific clinical data in a continuous mechanical model.
\newblock {\em PLoS ONE}, 10(7):e0132887, 2015.

\bibitem{Conte2020}
M.~Conte, L.~Gerardo-Giorda, and M.~Groppi.
\newblock Glioma invasion and its interplay with nervous tissue and therapy: A
  multiscale model.
\newblock {\em Journal of Theoretical Biology}, 486:110088, 2020.

\bibitem{CEKNSSW}
G.~Corbin, C.~Engwer, A.~Klar, J.~Nieto, J.~Soler, C.~Surulescu, and M.~Wenske.
\newblock Modeling glioma invasion with anisotropy- and hypoxia-triggered
  motility enhancement: from subcellular dynamics to macroscopic pdes with
  multiple taxis.
\newblock arXiv:2006.12322.

\bibitem{Corbin2018}
G.~Corbin, A.~Hunt, A.~Klar, F.~Schneider, and C.~Surulescu.
\newblock Higher-order models for glioma invasion: From a two-scale description
  to effective equations for mass density and momentum.
\newblock {\em Mathematical Models and Methods in Applied Sciences},
  28(09):1771--1800, 2018.

\bibitem{Czirok2013}
A.~Czirok.
\newblock Endothelial cell motility, coordination and pattern formation during
  vasculogenesis.
\newblock {\em Wiley Interdisciplinary Reviews: Systems Biology and Medicine},
  5(5):587--602, 2013.

\bibitem{diao2019}
W.~Diao, X.~Tong, C.~Yang, F.~Zhang, C.~Bao, H.~Chen, L.~Liu, M.~Li, F.~Ye,
  Q.~Fan, et~al.
\newblock Behaviors of glioblastoma cells in in vitro microenvironments.
\newblock {\em Scientific reports}, 9(1):1--9, 2019.

\bibitem{Engwer2}
C.~Engwer, T.~Hillen, M.~Knappitsch, and C.~Surulescu.
\newblock Glioma follow white matter tracts: a multiscale {DTI}-based model.
\newblock {\em Journal of Mathematical Biology}, 71(3):551--582, 2014.

\bibitem{EHS}
C.~Engwer, A.~Hunt, and C.~Surulescu.
\newblock Effective equations for anisotropic glioma spread with proliferation:
  a multiscale approach.
\newblock {\em IMA Journal of Mathematical Medicine and Biology}, 33:435--459,
  2016.

\bibitem{Engwer}
C.~Engwer, M.~Knappitsch, and C.~Surulescu.
\newblock A multiscale model for glioma spread including cell-tissue
  interactions and proliferation.
\newblock {\em Mathematical Biosciences and Engineering}, 13(2):443--460, 2016.

\bibitem{Engwer4}
C.~Engwer, C.~Stinner, and C.~Surulescu.
\newblock On a structured multiscale model for acid-mediated tumor invasion:
  The effects of adhesion and proliferation.
\newblock {\em Mathematical Models and Methods in Applied Sciences},
  27:1355--1390, 2017.

\bibitem{Gatenby}
R.A. Gatenby and E.T. Gawlinski.
\newblock A reaction-diffusion model of cancer invasion.
\newblock {\em Cancer Research}, 56(24):5745--5753, 1996.

\bibitem{Gatenby2006}
R.A. Gatenby, E.T. Gawlinski, A.F. Gmitro, B.~Kaylor, and R.J. Gillies.
\newblock Acid-mediated tumor invasion: a multidisciplinary study.
\newblock {\em Cancer Research}, 66(10):5216--5223, 2006.

\bibitem{Gerlee2012}
P.~Gerlee and S.~Nelander.
\newblock The impact of phenotypic switching on glioblastoma growth and
  invasion.
\newblock {\em {PLoS} Computational Biology}, 8(6):e1002556, 2012.

\bibitem{Giese2003}
A.~Giese, R.~Bjerkvig, M.E. Berens, and M.~Westphal.
\newblock Cost of migration: Invasion of malignant gliomas and implications for
  treatment.
\newblock {\em Journal of Clinical Oncology}, 21(8):1624--1636, 2003.

\bibitem{giese-etal96}
A.~Giese, L.~Kluwe, Meissner H., Michael E., and M.~Westphal.
\newblock Migration of human glioma cells on myelin.
\newblock {\em Neurosurgery}, 38:755--764, 1996.

\bibitem{Giese1996}
A.~Giese, M.A. Loo, N.~Tran, D.~Haskett, S.W. Coons, and M.E. Berens.
\newblock Dichotomy of astrocytoma migration and proliferation.
\newblock {\em International Journal of Cancer}, 67(2):275--282, 1996.

\bibitem{Hammoud1996}
M.A. Hammoud, R.~Sawaya, W.~Shi, P.F. Thall, and N.E. Leeds.
\newblock Prognostic significance of preoperative {MRI} scans in glioblastoma
  multiforme.
\newblock {\em Journal of Neuro-Oncology}, 27(1):65--73, 1996.

\bibitem{hanahan1996}
D.~Hanahan and J.~Folkman.
\newblock Patterns and emerging mechanisms of the angiogenic switch during
  tumorigenesis.
\newblock {\em Cell}, 86(3):353--364, 1996.

\bibitem{hanahan2011}
D.~Hanahan and R.A. Weinberg.
\newblock Hallmarks of cancer: the next generation.
\newblock {\em Cell}, 144(5):646--674, 2011.

\bibitem{Hardee2012}
M.E. Hardee and D.~Zagzag.
\newblock Mechanisms of glioma-associated neovascularization.
\newblock {\em The American Journal of Pathology}, 181(4):1126--1141, 2012.

\bibitem{Hatzikirou2010}
H.~Hatzikirou, D.~Basanta, M.~Simon, K.~Schaller, and A.~Deutsch.
\newblock Go or grow: the key to the emergence of invasion in tumour
  progression?
\newblock {\em Mathematical Medicine and Biology}, 29(1):49--65, 2010.

\bibitem{Hayat2011}
M.A. Hayat.
\newblock Introduction.
\newblock In {\em \emph{M.A. Hayat (ed.)}, Tumors of the Central Nervous
  System, Volume 1}, pages 3--8. Springer Netherlands, 2011.

\bibitem{VanderHeiden2009}
M.G.~Vander Heiden, L.C. Cantley, and C.B. Thompson.
\newblock Understanding the {Warburg} effect: The metabolic requirements of
  cell proliferation.
\newblock {\em Science}, 324(5930):1029--1033, 2009.

\bibitem{Henker2017}
C.~Henker, T.~Kriesen, \"{A}. Glass, B.~Schneider, and J.~Piek.
\newblock Volumetric quantification of glioblastoma: experiences with different
  measurement techniques and impact on survival.
\newblock {\em Journal of Neuro-Oncology}, 135(2):391--402, 2017.

\bibitem{Hinow}
P.~Hinow, P.~Gerlee, L.J. McCawley, V.~Quaranta, M.~Ciobanu, S.~Wang, J.M.
  Graham, B.P. Ayati, J.~Claridge, K.R. Swanson, M.~Loveless, and A.R.A.
  Anderson.
\newblock A spatial model of tumor-host interaction: application of
  chemotherapy.
\newblock {\em Mathematical Biosciences and Engineering}, 6(3):521--546, 2009.

\bibitem{Hiremath2015}
S.A. Hiremath and C.~Surulescu.
\newblock A stochastic multiscale model for acid mediated cancer invasion.
\newblock {\em Nonlinear Analysis: Real World Applications}, 22:176--205, 2015.

\bibitem{Athni_Hiremath_2016}
S.A. Hiremath and C.~Surulescu.
\newblock A stochastic model featuring acid-induced gaps during tumor
  progression.
\newblock {\em Nonlinearity}, 29(3):851--914, 2016.

\bibitem{HiSu-LNM}
S.A. Hiremath and C.~Surulescu.
\newblock Mathematical models for acid-mediated tumor invasion: from
  deterministic to stochastic approaches.
\newblock In {\em Multiscale models in mechano and tumor biology}, volume 122
  of {\em Lecture Notes in Computational Science and Engineering}, pages
  45--71. Springer, Cham, 2017.

\bibitem{AthniHiremath2018}
S.A. Hiremath, C.~Surulescu, A.~Zhigun, and S.~Sonner.
\newblock On a coupled {SDE}-{PDE} system modeling acid-mediated tumor
  invasion.
\newblock {\em Discrete {\&} Continuous Dynamical Systems - Series B},
  23(6):2339--2369, 2018.

\bibitem{Hogea2007}
C.~Hogea, C.~Davatzikos, and G.~Biros.
\newblock An image-driven parameter estimation problem for a
  reaction{\textendash}diffusion glioma growth model with mass effects.
\newblock {\em Journal of Mathematical Biology}, 56(6):793--825, 2007.

\bibitem{Holzer2009}
P.~Holzer.
\newblock Acid-sensitive ion channels and receptors.
\newblock In {\em Sensory Nerves}, pages 283--332. Springer Berlin Heidelberg,
  2009.

\bibitem{Hoering2012}
E.~H\"{o}ring, P.N. Harter, J.~Seznec, J.~Schittenhelm, H.-J. B\"{u}hring,
  S.~Bhattacharyya, E.~von Hattingen, C.~Zachskorn, M.~Mittelbronn, and
  U.~Naumann.
\newblock The {\textquotedblleft}go or grow{\textquotedblright} potential of
  gliomas is linked to the neuropeptide processing enzyme carboxypeptidase e
  and mediated by metabolic stress.
\newblock {\em Acta Neuropathologica}, 124(1):83--97, 2012.

\bibitem{Hunt2016}
A.~Hunt and C.~Surulescu.
\newblock A multiscale modeling approach to glioma invasion with therapy.
\newblock {\em Vietnam Journal of Mathematics}, 45(1-2):221--240, 2016.

\bibitem{Jbabdi2005}
S.~Jbabdi, E.~Mandonnet, H.~Duffau, L.~Capelle, K.R. Swanson,
  M.~P{\'{e}}l{\'{e}}grini-Issac, R.~Guillevin, and H.~Benali.
\newblock Simulation of anisotropic growth of low-grade gliomas using diffusion
  tensor imaging.
\newblock {\em Magnetic Resonance in Medicine}, 54(3):616--624, 2005.

\bibitem{Justus2013}
C.R. Justus, L.~Dong, and L.V. Yang.
\newblock Acidic tumor microenvironment and {pH}-sensing g protein-coupled
  receptors.
\newblock {\em Frontiers in Physiology}, 4, 2013.

\bibitem{Kaufman2005}
L.J. Kaufman, C.P. Brangwynne, K.E. Kasza, E.~Filippidi, V.D. Gordon, T.S.
  Deisboeck, and D.A Weitz.
\newblock Glioma expansion in collagen i matrices: analyzing collagen
  concentration-dependent growth and motility patterns.
\newblock {\em Biophysical journal}, 89(1):635--650, 2005.

\bibitem{ke2000}
L.D. Ke, Y.-X. Shi, S.-A. Im, X.~Chen, and W.K.A. Yung.
\newblock The relevance of cell proliferation, vascular endothelial growth
  factor, and basic fibroblast growth factor production to angiogenesis and
  tumorigenicity in human glioma cell lines.
\newblock {\em Clinical Cancer Research}, 6(6):2562--2572, 2000.

\bibitem{Kelkel2011}
J.~Kelkel and C.~Surulescu.
\newblock On some models for cancer cell migration through tissue networks.
\newblock {\em Mathematical Biosciences and Engineering}, 8(2):575--589, 2011.

\bibitem{KELKEL2012}
J.~Kelkel and C.~Surulescu.
\newblock A multiscale approach to cell migration in tissue networks.
\newblock {\em Mathematical Models and Methods in Applied Sciences},
  22(03):1150017, March 2012.

\bibitem{kim2009}
Y.~Kim, S.~Lawler, M.O. Nowicki, E.A. Chiocca, and A.~Friedman.
\newblock A mathematical model for pattern formation of glioma cells outside
  the tumor spheroid core.
\newblock {\em Journal of Theoretical Biology}, 260(3):359--371, 2009.

\bibitem{Kim2013}
Y.~Kim and S.~Roh.
\newblock A hybrid model for cell proliferation and migration in glioblastoma.
\newblock {\em Discrete \& Continuous Dynamical Systems - Series B},
  18(4):969--1015, 2013.

\bibitem{ko1980}
L~Ko, A~Koestner, and W~Wechsler.
\newblock Characterization of cell cycle and biological parameters of
  transplantable glioma cell lines and clones.
\newblock {\em Acta neuropathologica}, 51(2):107--111, 1980.

\bibitem{KSSSL}
N.~Kolbe, N.~Sfakianakis, C.~Stinner, C.~Surulescu, and J.~Lenz.
\newblock Modeling multiple taxis: tumor invasion with phenotypic
  heterogeneity, haptotaxis, and unilateral interspecies repellence.
\newblock arXiv:2005.01444v1.

\bibitem{Konukoglu2010}
E.~Konukoglu, O.~Clatz, P.-Y. Bondiau, H.~Delingette, and N.~Ayache.
\newblock Extrapolating glioma invasion margin in brain magnetic resonance
  images: Suggesting new irradiation margins.
\newblock {\em Medical Image Analysis}, 14(2):111--125, 2010.

\bibitem{Kumar20}
P.~Kumar, J.~Li, and C.~Surulescu.
\newblock Multiscale modeling of glioma pseudopalisades: contributions from the
  tumor microenvironment.
\newblock arXiv:2007.05297.

\bibitem{Lauffenburger}
D.A. Lauffenburger and J.L. Lindermann.
\newblock {\em Receptors. Models for binding, trafficing and signaling.}
\newblock Oxford University Press, 1993.

\bibitem{Lide}
D.R. Lide~(ed).
\newblock {\em CRC handbook of chemistry and physics, vol. 85}.
\newblock CRC Press, 2004.

\bibitem{Lorenz2014}
T.~Lorenz and C.~Surulescu.
\newblock On a class of multiscale cancer cell migration models: Well-posedness
  in less regular function spaces.
\newblock {\em Mathematical Models and Methods in Applied Sciences},
  24(12):2383--2436, 2014.

\bibitem{lunt}
S.Y. Lunt and M.G. Vander~Heiden.
\newblock Aerobic glycolysis: Meeting the metabolic requirements of cell
  proliferation.
\newblock {\em Annual Review of Cell and Developmental Biology},
  27(1):441--464, 2011.
\newblock PMID: 21985671.

\bibitem{Martin2}
G.R. Martin and R.K. Jain.
\newblock Noninvasive measurement of interstitial ph profiles in normal and
  neoplastic tissue using fluorescence ratio imaging microscopy.
\newblock {\em Cancer Research}, 54(21):5670--5674, 1994.

\bibitem{Martin}
N.K. Martin, E.A. Gaffney, R.A. Gatenby, R.J. Gillies, I.F. Robey, and P.K.
  Maini.
\newblock A mathematical model of tumour and blood {pHe} regulation: The
  buffering system.
\newblock {\em Mathematical Biosciences}, 230(1):1--11, 2011.

\bibitem{MartnezGonzlez2012}
A.~Mart{\'{\i}}nez-Gonz{\'{a}}lez, G.F. Calvo, L.A.~P{\'{e}}rez Romasanta, and
  V.M. P{\'{e}}rez-Garc{\'{\i}}a.
\newblock Hypoxic cell waves around necrotic cores in glioblastoma: A
  biomathematical model and its therapeutic implications.
\newblock {\em Bulletin of Mathematical Biology}, 74(12):2875--2896, 2012.

\bibitem{MartnezZaguiln1996}
R.~Mart\'inez-Zaguil\'an, E.A. Seftor, R.E.B. Seftor, Y.-W. Chu, R.J. Gillies,
  and M.J.C. Hendrix.
\newblock Acidic {pH} enhances the invasive behavior of human melanoma cells.
\newblock {\em Clinical {\&} Experimental Metastasis}, 14(2):176--186, 1996.

\bibitem{McGillen2013}
J.B. McGillen, E.A. Gaffney, N.K. Martin, and P.K. Maini.
\newblock A general reaction{\textendash}diffusion model of acidity in cancer
  invasion.
\newblock {\em Journal of Mathematical Biology}, 68(5):1199--1224, 2013.

\bibitem{Meral2015-JCSMS}
G.~Meral, C.~Stinner, and C.~Surulescu.
\newblock A multiscale model for acid-mediated tumor invasion: Therapy
  approaches.
\newblock {\em Journal of Coupled Systems and Multiscale Dynamics},
  3(2):135--142, 2015.

\bibitem{Moellering2008}
R.E. Moellering, K.C. Black, C.~Krishnamurty, B.K. Baggett, P.~Stafford,
  M.~Rain, R.A. Gatenby, and R.J. Gillies.
\newblock Acid treatment of melanoma cells selects for invasive phenotypes.
\newblock {\em Clinical {\&} Experimental Metastasis}, 25(4):411--425, 2008.

\bibitem{Murray1989}
J.D. Murray.
\newblock {\em Mathematical Biology}.
\newblock Springer Berlin Heidelberg, 1989.

\bibitem{Othmer2002}
H.G. Othmer and T.~Hillen.
\newblock The diffusion limit of transport equations {II}: Chemotaxis
  equations.
\newblock {\em {SIAM} Journal on Applied Mathematics}, 62(4):1222--1250, 2002.

\bibitem{PH13}
K.J. Painter and T.~Hillen.
\newblock Mathematical modelling of glioma growth: the use of diffusion tensor
  imaging (dti) data to predict the anisotropic pathways of cancer invasion.
\newblock {\em Journal of Theoretical Biology}, 323:25--39, 2013.

\bibitem{pennarun2005}
G.~Pennarun, C.~Granotier, L.R. Gauthier, D.~Gomez, F.~Hoffschir, E.~Mandine,
  J.-F. Riou, J.-L. Mergny, P.~Mailliet, and F.D. Boussin.
\newblock Apoptosis related to telomere instability and cell cycle alterations
  in human glioma cells treated by new highly selective g-quadruplex ligands.
\newblock {\em Oncogene}, 24(18):2917--2928, 2005.

\bibitem{Perthame2016}
B.~Perthame, M.~Tang, and N.~Vauchelet.
\newblock Derivation of the bacterial run-and-tumble kinetic equation from a
  model with biochemical pathway.
\newblock {\em Journal of Mathematical Biology}, 73(5):1161--1178, 2016.

\bibitem{Pham2012}
K.~Pham, A.~Chauviere, H.~Hatzikirou, X.~Li, H.M. Byrne, V.~Cristini, and
  J.~Lowengrub.
\newblock Density-dependent quiescence in glioma invasion: instability in a
  simple reaction{\textendash}diffusion model for the migration/proliferation
  dichotomy.
\newblock {\em Journal of Biological Dynamics}, 6(sup1):54--71, 2012.

\bibitem{Pignatti2002}
F.~Pignatti, M.~van~den Bent, D.~Curran, C.~Debruyne, R.~Sylvester,
  P.~Therasse, D.~{\'{A}}fra, P.~Cornu, M.~Bolla, C.~Vecht, and A.B.M.F. Karim.
\newblock Prognostic factors for survival in adult patients with cerebral
  low-grade glioma.
\newblock {\em Journal of Clinical Oncology}, 20(8):2076--2084, 2002.

\bibitem{Prag}
S.~Prag, E.A. Lepekhin, K.~Kolkova, R.~Hartmann-Petersen, A.~Kawa, P.S. Walmod,
  V.~Belman, H.C. Gallagher, V.~Berezin, E.~Bock, and N.~Pedersen.
\newblock Ncam regulates cell motility.
\newblock {\em Journal of Cell Science}, 115(2):283--292, 2002.

\bibitem{rojiani1996}
A.M. Rojiani and K.~Dorovini-Zis.
\newblock Glomeruloid vascular structures in glioblastoma multiforme: an
  immunohistochemical and ultrastructural study.
\newblock {\em Journal of neurosurgery}, 85(6):1078--1084, 1996.

\bibitem{Semenza2009}
G.L. Semenza.
\newblock Defining the role of hypoxia-inducible factor 1 in cancer biology and
  therapeutics.
\newblock {\em Oncogene}, 29(5):625--634, 2009.

\bibitem{Sidani}
M.~Sidani, D.~Wessels, G.~Mouneimne, M.~Ghosh, S.~Goswami, C.~Sarmiento,
  W.~Wang, S.~Kuhl, M.~El-Sibai, J-M. Backer, R.~Eddy, D.~Soll, and
  J.~Condeelis.
\newblock Cofilin determines the migration behavior and turning frequency of
  metastatic cancer cells.
\newblock {\em Journal of Cell Biology}, 179(4):777--791, 2007.

\bibitem{Smallbone2008}
K.~Smallbone, R.A. Gatenby, and P.K. Maini.
\newblock Mathematical modelling of tumour acidity.
\newblock {\em Journal of Theoretical Biology}, 255(1):106--112, 2008.

\bibitem{stinner-surulescu-meral}
C.~Stinner, C.~Surulescu, and G.~Meral.
\newblock A multiscale model for {pH}-tactic invasion with time-varying
  carrying capacities.
\newblock {\em IMA Journal of Applied Mathematics}, 80:1300--1321, 2015.

\bibitem{stinner-surulescu-uatay}
C.~Stinner, C.~Surulescu, and A.~Uatay.
\newblock Global existence for a go-or-grow multiscale model for tumor invasion
  with therapy.
\newblock {\em Mathematical Models and Methods in Applied Sciences},
  26:2163--2201, 2016.

\bibitem{Stokes1991}
C.L. Stokes and D.A. Lauffenburger.
\newblock Analysis of the roles of microvessel endothelial cell random motility
  and chemotaxis in angiogenesis.
\newblock {\em Journal of Theoretical Biology}, 152(3):377--403, 1991.

\bibitem{Stokes1990}
C.L. Stokes, M.A. Rupnick, S.K. Williams, and D.A. Lauffenburger.
\newblock Chemotaxis of human microvessel endothelial cells in response to
  acidic fibroblast growth factor.
\newblock {\em Laboratory investigation; a journal of technical methods and
  pathology}, 63(5):657—668, 1990.

\bibitem{Swan2017}
A.~Swan, T.~Hillen, J.C. Bowman, and A.D. Murtha.
\newblock A patient-specific anisotropic diffusion model for brain tumour
  spread.
\newblock {\em Bulletin of Mathematical Biology}, 80(5):1259--1291, 2017.

\bibitem{Swanson2011}
K.R. Swanson, R.C. Rockne, J.~Claridge, M.A. Chaplain, E.C. Alvord, and A.R.A.
  Anderson.
\newblock Quantifying the role of angiogenesis in malignant progression of
  gliomas: In silico modeling integrates imaging and histology.
\newblock {\em Cancer Research}, 71(24):7366--7375, 2011.

\bibitem{Szabo2010}
A.~Szabó, R.~Ünnep, E.~Méhes, W.~O. Twal, W.~S. Argraves, Y.~Cao, and
  A~Czirók.
\newblock Collective cell motion in endothelial monolayers.
\newblock {\em Physical Biology}, 7(4):046007, 2010.

\bibitem{Vaupel}
P.~Vaupel, F.~Kallinowski, and P.~Okunieff.
\newblock Blood flow, oxygen and nutrient supply, and metabolic
  microenvironment of human tumors: A review.
\newblock {\em Cancer Research}, 49(23):6449--6465, 1989.

\bibitem{Wang2019}
Y.-L. Wang, J.~Yao, A.~Chakhoyan, C.~Raymond, N.~Salamon, L.M. Liau, P.L.
  Nghiemphu, A.~Lai, W.B. Pope, N.~Nguyen, M.~Ji, T.F. Cloughesy, and B.M.
  Ellingson.
\newblock Association between tumor acidity and hypervascularity in human
  gliomas using {pH}-weighted amine chemical exchange saturation transfer
  echo-planar imaging and dynamic susceptibility contrast perfusion {MRI} at
  3t.
\newblock {\em American Journal of Neuroradiology}, 40(6):979--986, 2019.

\bibitem{Webb2011}
B.A. Webb, M.~Chimenti, M.P. Jacobson, and D.L. Barber.
\newblock Dysregulated {pH}: a perfect storm for cancer progression.
\newblock {\em Nature Reviews Cancer}, 11(9):671--677, 2011.

\bibitem{Weisz2017}
K.T. Wei{\ss}, M.~Fante, G.~K\"{o}hl, J.~Schreml, F.~Haubner, M.~Kreutz,
  S.~Haverkampf, M.~Berneburg, and S.~Schreml.
\newblock Proton-sensing g protein-coupled receptors as regulators of cell
  proliferation and migration during tumor growth and wound healing.
\newblock {\em Experimental Dermatology}, 26(2):127--132, 2017.

\bibitem{Wippold2037}
F.J. Wippold, M.~L{\"a}mmle, F.~Anatelli, J.~Lennerz, and A.~Perry.
\newblock Neuropathology for the neuroradiologist: Palisades and
  pseudopalisades.
\newblock {\em American Journal of Neuroradiology}, 27(10):2037--2041, 2006.

\bibitem{wrensch}
M.~Wrensch, Y.~Minn, T.~Chew, M.~Bondy, and M.S. Berger.
\newblock Epidemiology of primary brain tumors: Current concepts and review of
  the literature.
\newblock {\em Neuro-Oncology}, 4(4):278--299, 2002.

\bibitem{Xie2014}
Q.~Xie, S.~Mittal, and M.~E. Berens.
\newblock Targeting adaptive glioblastoma: an overview of proliferation and
  invasion.
\newblock {\em Neuro-Oncology}, 16(12):1575--1584, 2014.

\bibitem{Xu2001}
L.~Xu, D.~Fukumura, and R.K. Jain.
\newblock Acidic extracellular {pH} induces vascular endothelial growth factor
  ({VEGF}) in human glioblastoma cells via {ERK}1/2 {MAPK} signaling pathway.
\newblock {\em Journal of Biological Chemistry}, 277(13):11368--11374, 2001.

\bibitem{Zhigun2018}
A.~Zhigun, C.~Surulescu, and A.~Hunt.
\newblock A strongly degenerate diffusion-haptotaxis model of tumour invasion
  under the go-or-grow dichotomy hypothesis.
\newblock {\em Mathematical Methods in the Applied Sciences}, 41:2403--2428,
  2018.

\end{thebibliography}

\end{document}